\documentclass[aps, prx,twocolumn,longbibliography,superscriptaddress,amsmath,amssymb,floatfix]{revtex4}
\usepackage[latin9]{inputenc}
\usepackage{amssymb}
\usepackage{graphicx}
\usepackage{amsmath}
\usepackage{color}
\usepackage{mathrsfs}
\usepackage{float}
\usepackage{indentfirst}
\usepackage{mathrsfs}
\usepackage{float}
\usepackage{indentfirst}
\usepackage{textcomp}
\usepackage{comment}
\usepackage{mathtools}
\usepackage{natbib,hyperref}
\usepackage{soul}

\begin{document}

\title{Stripe order in the doped Hubbard model on the honeycomb lattice}

\author{Xu Yang}
\affiliation{Key Laboratory of Artificial Structures and Quantum Control (Ministry of Education),  School of Physics and Astronomy, Shanghai Jiao Tong University, Shanghai 200240, China}
\affiliation{Shenyang National Laboratory for Materials Science, Shanghai Jiao Tong University,
	800 Dongchuan Road, Shanghai 200240, China}
\author{Hao Zheng} \affiliation{Key Laboratory of Artificial Structures and Quantum Control (Ministry of Education),  School of Physics and Astronomy, Shanghai Jiao Tong University, Shanghai 200240, China}
\affiliation{Shenyang National Laboratory for Materials Science, Shanghai Jiao Tong University,
	800 Dongchuan Road, Shanghai 200240, China}
\author{Mingpu Qin} \thanks{qinmingpu@sjtu.edu.cn}
\affiliation{Key Laboratory of Artificial Structures and Quantum Control (Ministry of Education),  School of Physics and Astronomy, Shanghai Jiao Tong University, Shanghai 200240, China}

\begin{abstract}
We study the ground state properties of the doped Hubbard model with strong interactions on honeycomb lattice by the Density Matrix Renormalization Group (DMRG)
method. At half-filling, due to the absence of minus sign problem, it is now well established by large-scale Quantum Monte Carlo calculations that a
Dirac semi-metal to anti-ferromagnetic Mott insulator transition occurs with the increase of the interaction
strength $U$ for the Hubbard model on honeycomb lattice. However, an understanding of the fate of the anti-ferromagnetic Mott insulator when holes are doped into the system is
still lacking. In this work, by calculating the local spin and charge density for width-4 cylinders with DMRG, we discover a half-filled stripe order in the doped Hubbard model 
on honeycomb lattice. We also perform complementary large-scale mean-field calculations with renormalized interaction strength. We observe half-filled stripe order and
find stripe states with filling close to one half are nearly degenerate in energy.      

\end{abstract}

\maketitle
 
\section{Introduction}

The Hubbard model is one of the fundamental model systems in the exploration of quantum many-body
effect \cite{Hubbard238}. It captures many important aspects of correlated electrons by considering only
the on-site Coulomb interactions.
The Hubbard model on square lattice is related to the copper oxide superconductors \cite{PhysRevB.37.3759,ANDERSON1196,RevModPhys.66.763,RevModPhys.84.1383}
and has been extensively studied numerically \cite{PhysRevX.5.041041}.
The Hubbard model on honeycomb lattice, which has a similar structure as square lattice, is also widely studied, partly for its
connection to graphene \cite{RevModPhys.81.109}. It is also a playing ground to study the correlation-driven metal-insulator
transition \cite{RevModPhys.70.1039}.
The properties of the Hubbard model on honeycomb lattice at half-filling can be accurately
determined by Quantum Monte Carlo (QMC) method due to the absence of the minus-sign problem on bipartite lattices \cite{Sorella_1992}.
 The ground state phase diagram at half-filling is now well
established with large-scale QMC calculations \cite{Sorella_1992,Sorella_2012,PhysRevX.3.031010}. At half-filling, in contrast to the Hubbard model
on square lattice where there is no metal-insulator transition at finite interaction strength \cite{PhysRevB.94.085140}, a transition
occurs from the Dirac semi-metal phase
at weak interactions to the Mott insulator phase at strong interactions on the honeycomb lattice. 
The critical interaction strength is estimated to be $U_c \approx 3.8$ \cite{PhysRevX.3.031010,PhysRevX.6.011029} by large-scale QMC calculations
with careful finite-size analysis.
The transition is found to be in the Gross-Neveu-Yukawa \cite{PhysRevX.3.031010,PhysRevX.6.011029} universality class. Anti-ferromagnetic (AF)
long-range order is also found to develop accompanying the Mott transition \cite{Sorella_1992,Sorella_2012,PhysRevX.3.031010}. In the large interaction
strength limit at
half-filling, the effective low energy Heisenberg model on honeycomb lattice is known to order anti-ferromanetically with local momentum $m = 0.2677(6)$ \cite{Reger_1989,PhysRevB.73.054422}. 

The fate of AF Mott insulator if holes are doped in the Hubbard model on honeycomb lattice is still unknown. Away from half-filling, the infamous minus sign problem
emerges which hampers the application of QMC to large system sizes and low temperature \cite{PhysRevB.41.9301,PhysRevLett.94.170201}. The competition between kinetic and
potential energies can lead to exotic states when holes are introduced into the Mott insulator \cite{ANDERSON1196}. In cuprates, where the
CuO plane has a square lattice structure, high-$T_c$ superconductivity can emerge by doping holes into the parent antiferromagnetic Mott insulator \cite{Bednorz1986}. 
Stripe order \cite{nature_375_15_1995} is also observed in the phase diagram of cuprates. 
For the Hubbard model on square lattice, which is believed to be relevant to cuprates,
different ground states are obtained \cite{PhysRevLett.91.136403,PhysRevB.71.075108,PhysRevLett.95.237001}.
A recent work by a collaboration of state-of-art numerical approaches \cite{Zheng1155}
shows the ground state of the Hubbard model on square lattice has stripe order in the under-doped region
with strong interactions.
Results also indicate stripe order competes with possible d-wave superconducting order on square lattice \cite{PhysRevX.10.031016}.
 Given the similarity of honeycomb and square lattice, e.g., both of
them are bipartite and order anti-ferromagnetically in the strong interaction region at half filling, 
it is natural to ask whether stripe order also exists
on the honeycomb lattice when holes are doped into the AF Mott insulator phase.  

Many attempts have been made trying to reveal the properties of the doped Hubbard model on honeycomb lattice.
Substantial attentions have been paid to the $1/4$ doping case, where the density of states displays a Van Hove singularity and 
the Fermi surface has a nesting feature.
At weak interaction, which is relevant to graphene \cite{RevModPhys.81.109}, $d+id$ superconductivity was found in the Hubbard model on honeycomb lattice near $1/4$ doping
by variational Monte Carlo \cite{PhysRevB.81.085431},
renormalization group \cite{PhysRevLett.100.146404,Nandkishore_2012,PhysRevB.86.020507,PhysRevB.81.224505}, singular-mode functional renormalization
group and variational Monte Carlo \cite{PhysRevB.85.035414}, and by a combination of different numerical methods \cite{PhysRevX.4.031040}.
Spontaneous quantum Hall effect is also found at $1/4$ doping \cite{Li_2012,PhysRevB.85.035414,PhysRevX.4.031040}.  
In a recent work, $p+ip$ superconductivity was obtained with Grassmann tensor product state approach \cite{PhysRevB.101.205147} in the infinite-U limit, i.e., the t-J model. 
A detailed analysis of the possible pairing symmetry can be found in \cite{PhysRevB.90.054521}. 

Experimentally, real materials with honeycomb structure other than graphene \cite{RevModPhys.81.109} were also synthesized. Long-range AF Neel order was observed in real materials with honeycomb structure, ${\text{Na}}_{2}{\text{IrO}}_{3}$ \cite{PhysRevB.82.064412}
and $\text{InCu}_{\frac{2}{3}}\text{V}_{\frac{1}{3}}\text{O}_3$ \cite{KATAEV2005310} for examples. Superconductivity was also discovered in the pnictide SrPtAs which
has a honeycomb structure \cite{doi:10.1143/JPSJ.80.055002}. Time-reversal symmetry was also found to be broken in the superconducting
state of SrPtAs \cite{PhysRevB.87.180503,PhysRevB.89.020509}. 
A pressure-driven superconductivity in $\text{FeP}\text{Se}_3$, which has an iron-based honeycomb lattice structure was reported recently \cite{Wang18}.      

In this work we study the ground state properties of the doped Hubbard model on honeycomb lattice in the strong interaction region, i.e., in the AF Mott 
insulator phase. We employ
the Density Matrix Renormalization Group (DMRG) method with which reliable results of ground state are obtained.
We discover a half-filled stripe order at $1/16$ doping, which is similar as the stripe state on the square lattice \cite{2020arXiv200910736W}. We also perform
large-scale mean-field calculations with renormalized interaction strength for large systems. We find half-filled stripe state in the mean-field calculations. Stripe
state with filling other one half are also obtained and the energies for stripe states near half filling are almost degenerate in mean-field level which indicates the wave-length of
stripe can fluctuate without causing much energy, a phenomenon also observed in the square lattice case \cite{Zheng1155}. 

\begin{figure}[t]
	\includegraphics[width=40mm]{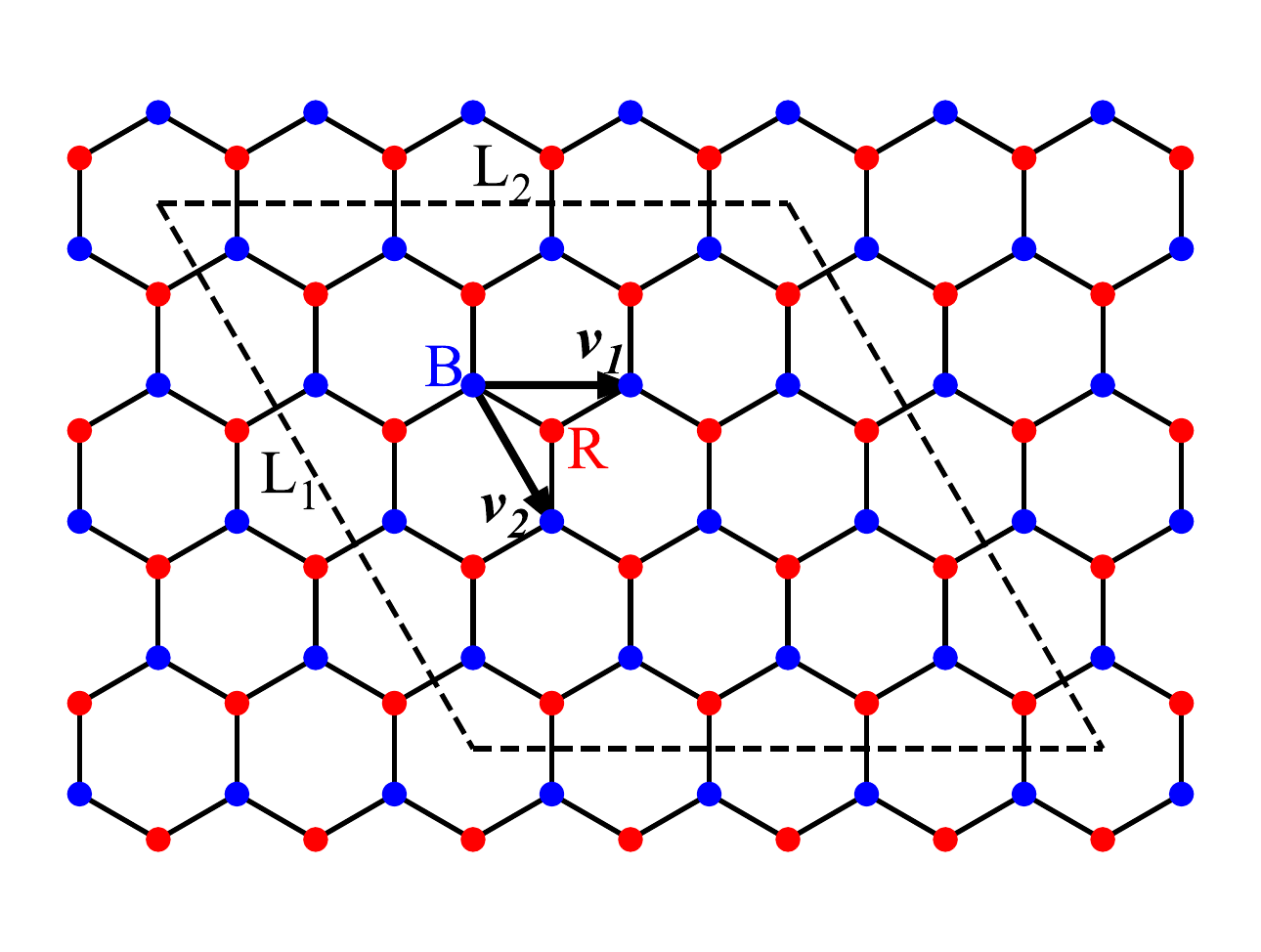}
	\includegraphics[width=40mm]{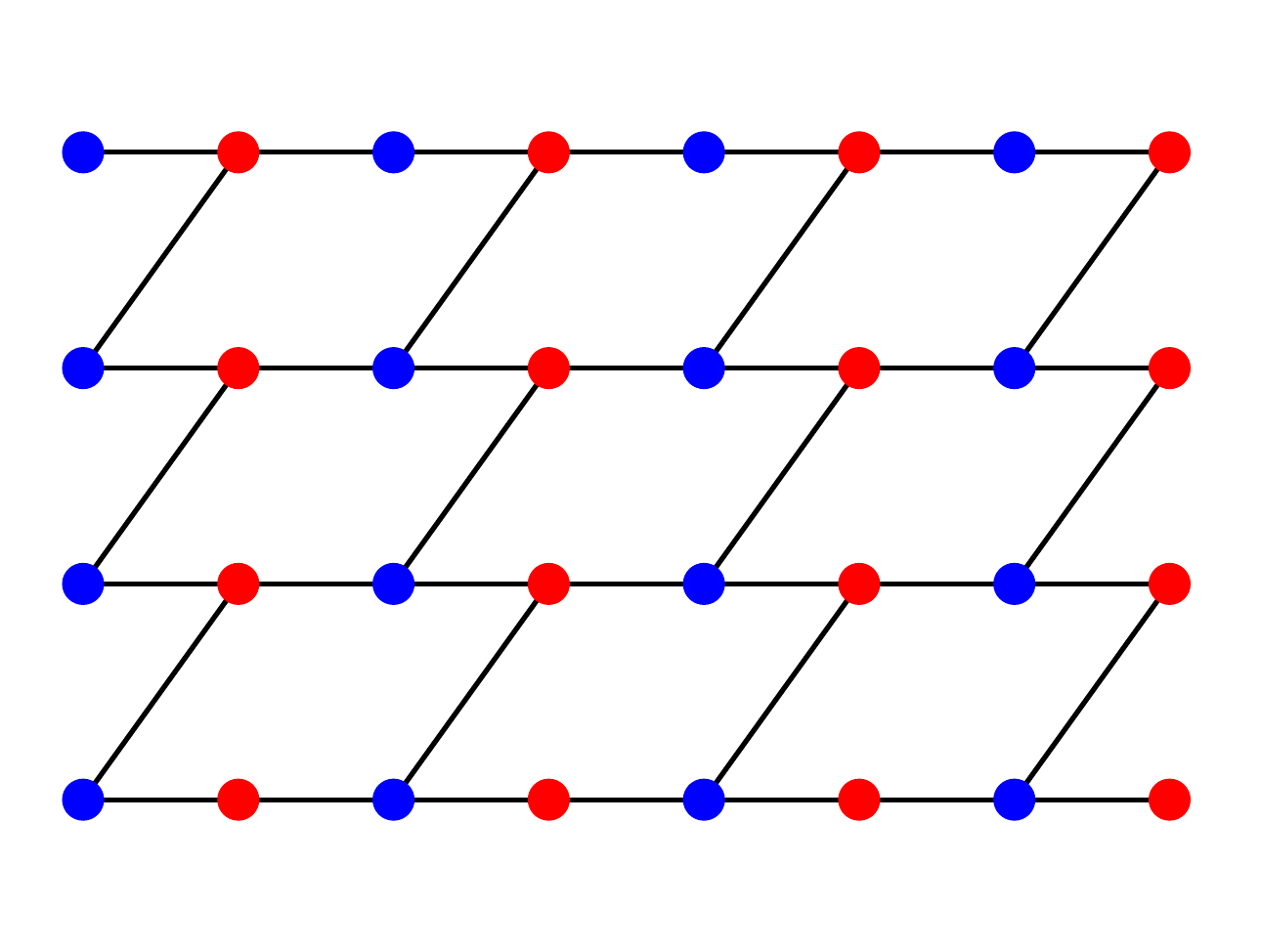}
	\caption{Left: sketch of the honeycomb lattice. Blue and red dots represent sites for the two sub-lattices respectively. We choose the two primitive vectors as \bm{$v_{1}$} and \bm{$v_{2}$}. The dotted rhombus contains a $L_{1} \times L_{2} = 4 \times 4$ super-cell. Right: in DMRG calculation, we arrange the super-cell into a square lattice which is then mapped into a one-dimensional chain in the ordinary manner. The $4 \times 4$ cell in the left panel is transformed into a square lattice with size $4 \times 8$.} 
	\label{hexa}
\end{figure}

\section{Model and Methods}

The Hamiltonian of Hubbard model is:
\begin{equation}
H=-t\sum_{\langle i,j\rangle,\sigma}c_{i\sigma}^\dagger c_{j\sigma}+U\sum_{i}n_{i\uparrow}n_{i\downarrow}
\end{equation}
where $c_{i,\sigma}$, $c_{i,\sigma}^{\dagger}$ and $n_{i\sigma}=c_{i\sigma}^\dagger c_{i\sigma}$ are the
annihilation, creation and density operators on site $i$ for spin species $\sigma$.   
$\langle i,j \rangle$ denotes the nearest neighboring hopping on the honeycomb lattice. $U > 0$ is the repulsive interaction strength and $t$ is
the hopping constant which is set to the energy unit. The total number of electrons with spin $\sigma(\sigma=\uparrow,\downarrow)$ are denoted by $N_{\sigma}$. 
We only consider the spin-balanced case so the total number of electrons is $N_e = 2 N_{\sigma}$. We 
denote the number of primitive cells in the studied system as $N = L_1 \times L_2$
(see the left panel of Fig.~\ref{hexa}), so the total number of sites is $2N$ because each primitive cell contains two sites. The doping level is $h = \frac{N_{hole}}{2N}$.
where $N_{hole}=2N-N_e$ is the total number of holes. The local spin and hole density at site $i$ is $S_{i} = (n_{i,\uparrow}-n_{i, \downarrow}) / 2$ and
$h_i = (1 - n_{i,\uparrow}-n_{i, \downarrow})$ respectively. 
In DMRG calculations, we adopt cylinder geometry, i.e., periodic (open) boundary conditions along $L_1$($L_2$) directions, while periodic
	boundary conditions in both directions are used in mean-field calculations \cite{foot1}. 

To characterize the stripe structure, staggered spin density which is defined as $(-1)^iS_i$ is
plotted.

\begin{figure}[t]
	\includegraphics[width=80mm]{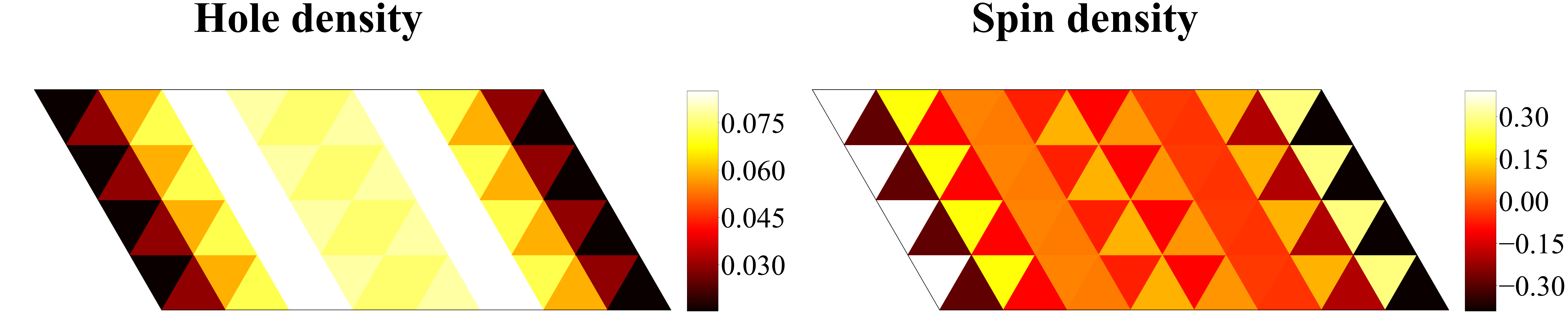}
	\includegraphics[width=80mm]{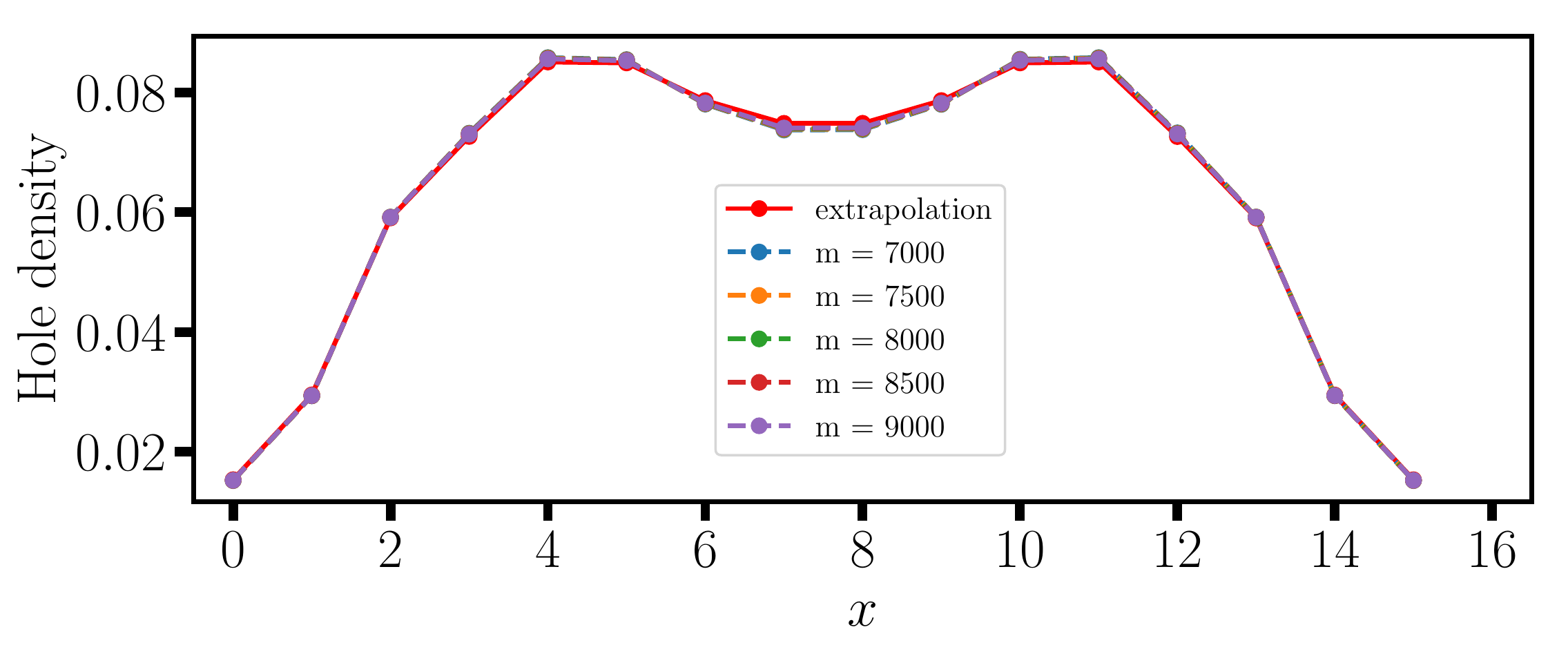}
	\includegraphics[width=80mm]{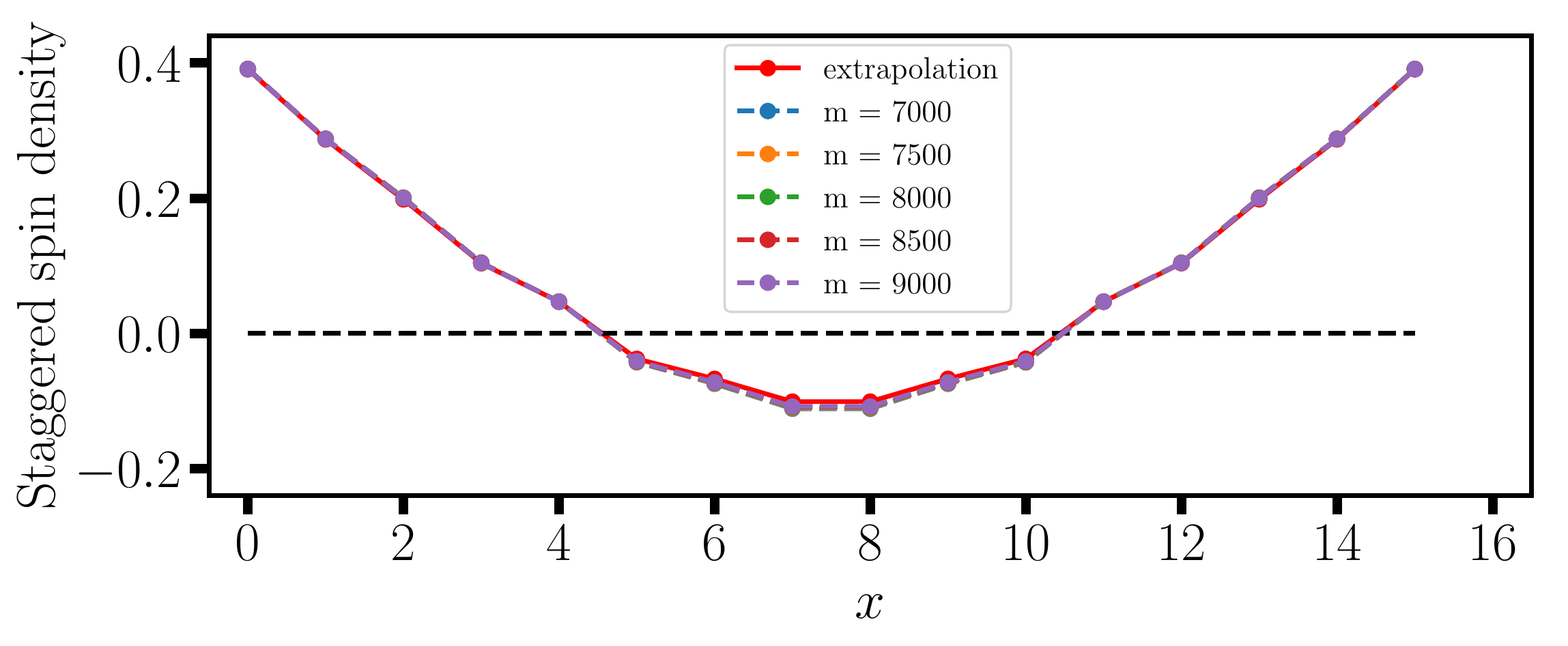}
	\caption{DMRG results of the hole and spin density for a $4 \times 8$ cylinder with $U = 8$ and $4$ holes.
		Top: color map of the hole (left) and spin density (right). Middle (Bottom): plots of hole (staggered spin) density along $L_{2}$ direction.
		Both results with finite kept state $m$ and the extrapolated to zero truncation error results are shown. The details of the
		extrapolation can be found in the Appendix \cite{supp}.}
	\label{DMRG_4_8}
\end{figure}

DMRG \cite{PhysRevLett.69.2863,PhysRevB.48.10345} is an extremely 
accurate method and arguably the workhorse for the study of one-dimensional chains and narrow cylinders \cite{RevModPhys.77.259,doi:10.1146/annurev-conmatphys-020911-125018}.
In DMRG calculation, we focus on $1/16$ doping and $U = 8$. We calculate $4 \times 8$ and $4 \times 16$ systems which contain $64$ and $128$ sites
respectively. The state kept in
the DMRG calculation is as large as $m = 9000$ to make sure the linear scaling region is reached and linear extrapolations with truncation error for physical quantities are 
performed.

To complement the DMRG study of width-4 cylinders, we also perform large-scale mean-field calculation on systems with larger size. It is known that
mean-field calculation usually exaggerates the order parameter \cite{PhysRevB.94.235119}, so in the mean-field calculation we choose a smaller
renormalized interaction strength $U \approx 3.0$ (above the critical interaction strength $U_c^{mf} \approx 2.23 $ in the mean-field phase
diagram \cite{PhysRevB.101.125103}). Details of the mean-field calculations can be found in the Appendix \cite{supp}.

\section{Results}

\begin{figure}[t]
	\includegraphics[width=80mm]{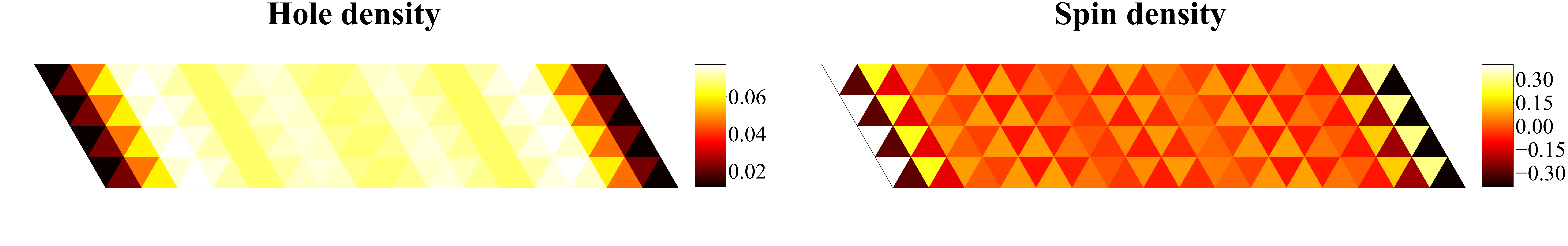}
	\includegraphics[width=80mm]{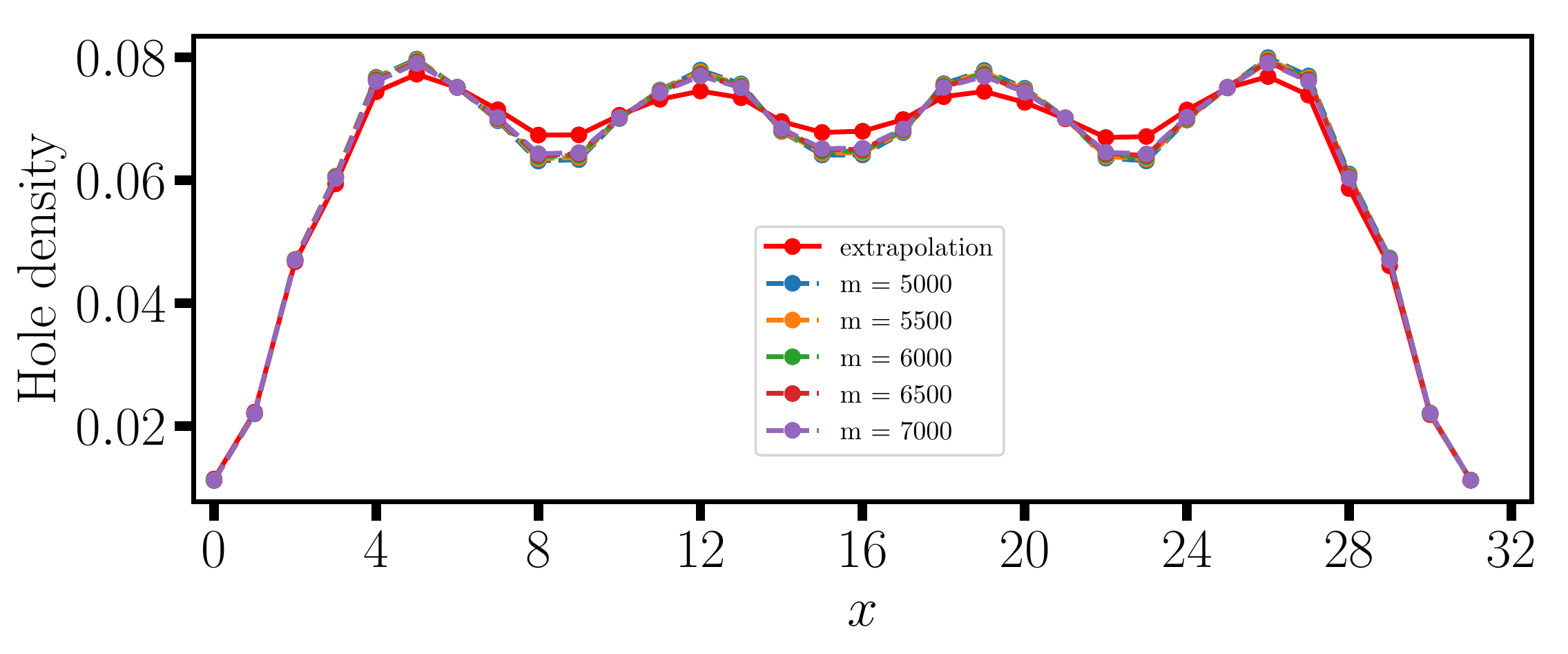}
	\includegraphics[width=80mm]{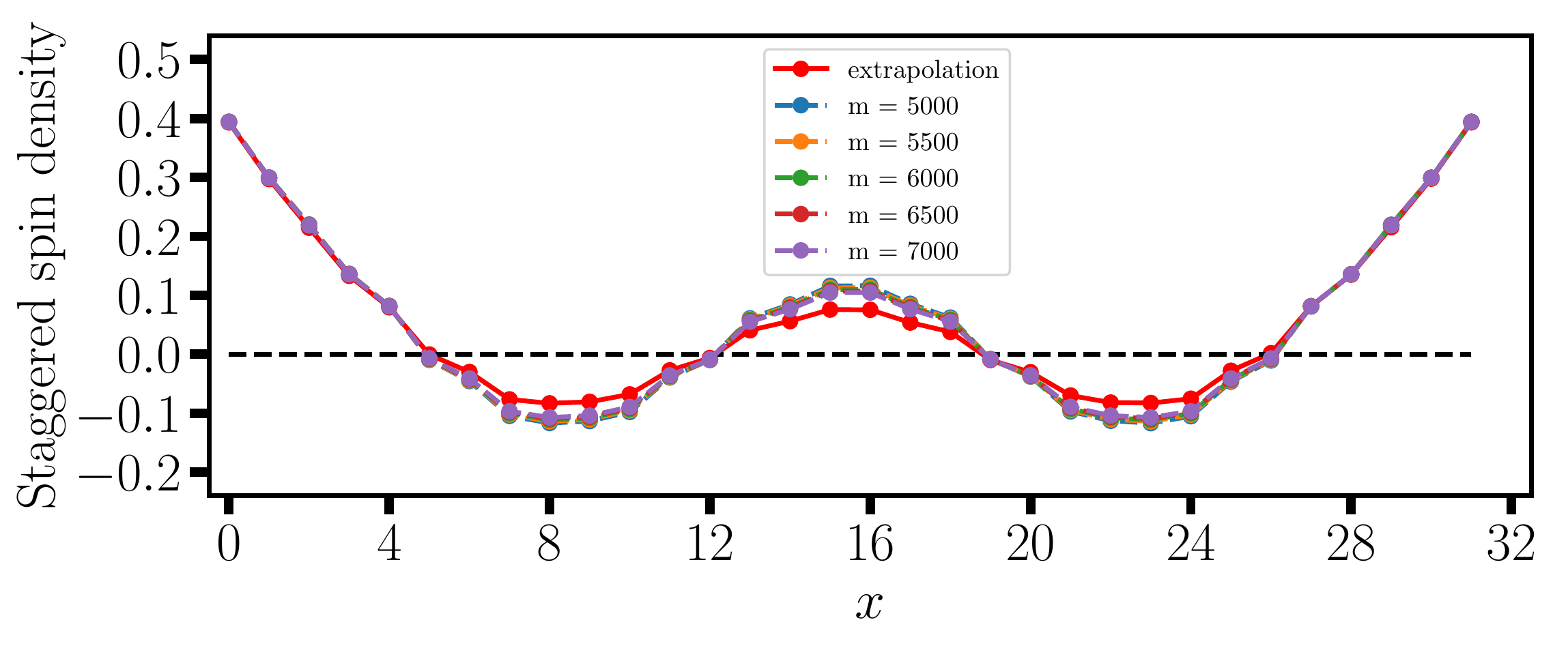}
	\caption{Similar as Fig.~\ref{DMRG_4_8}. The system is a $4 \times 16$ cylinder with $U = 8$ and $8$ holes.}
	\label{DMRG_4_16}
\end{figure}

\subsection{DMRG}
 In DMRG calculation, we rearrange the honeycomb lattice into a square lattice with next nearest neighboring interactions (see the right
panel of Fig.~\ref{hexa}) 
and map the square lattice to a one-dimensional chain in a common manner \cite{PhysRevB.95.125125}.
In Fig.~\ref{DMRG_4_8} we plot the local hole and staggered spin density for the $4 \times 8$ system at $1/16$ doping 
with $U = 8$. We apply anti-ferromagnetic pinning fields with strength $h_p = 0.5$ on the open edges of the cylinder to break the SU(2) symmetry explicitly.
In this way, we can measure the local spin density to probe the possible spin order instead of the spin-spin correlation function, which is more difficult to
calculate \cite{PhysRevLett.99.127004}.
Color maps and plots for
both the hole and staggered density along $L_{2}$ direction are shown in Fig.~\ref{DMRG_4_8}. DMRG results with both finite kept state
 $m$ (from $7000$ to $9000$) and the extrapolated
to zero truncation error results are presented \cite{PhysRevLett.99.127004}. In the Appendix \cite{supp}, we show the
scaling of energy and hole density with truncation error and find both of them enter the linear scaling region which ensures the reliability of the
extrapolation.   

In Fig.~\ref{DMRG_4_16} we study a longer cylinder with size $4 \times 16$ and $U = 8$ also at $1/16$ doping. Same as in Fig.~\ref{DMRG_4_8}, color map and plots for
both the hole and staggered density along $L_{2}$ direction are shown. 

From both Fig.~\ref{DMRG_4_8} and Fig.~\ref{DMRG_4_16}, we can find a periodic oscillation of both the hole and staggered spin density.
The wave-length of staggered spin density is twice that of hole density. At the peak position of hole density,
where holes are concentrated, the spin density shows a $\pi$ phase shift which is the feature of the stripe order \cite{nature_375_15_1995}. The stripe order is half-filled
because there are $2 (4)$ stripes and $4 (8)$ holes in Fig.~\ref{DMRG_4_8} (Fig.~\ref{DMRG_4_16}), and the width of the system is $4$. 
The amplitude of hole modulation on honeycomb lattice is smaller than that on the square lattice \cite{PhysRevB.94.235119,Zheng1155} because of the larger quantum fluctuation on
honeycomb lattice \cite{foot2}. 

The fast increase of the number of required kept states in DMRG with width of the system prevents us to reach cylinder beyond width $4$.
In general, the quantum fluctuation is larger for narrow cylinders as shown in the evolution of AF order in Heisenberg model from
one to two dimension. The stripe order was found to increase in wider cylinders on square lattice \cite{PhysRevB.94.235119,Zheng1155}.  
Nevertheless, we explore the boundary (pinning field) effect in the Appendix \cite{supp}. We find the stripe order in the bulk of the system is robust against the decrease of the strength of pinning fields.

It is worth noting that the stripe order in honeycomb lattice is in the diagonal direction of the underlying square lattice if we arrange the honeycomb
lattice in the brick-wall way. The diagonal stripe state was found to be
in the low energy manifold of the $t-J$ model on square lattice \cite{PhysRevLett.113.046402} with tensor network states method. 


\begin{figure}[t]
	\includegraphics[width=80mm]{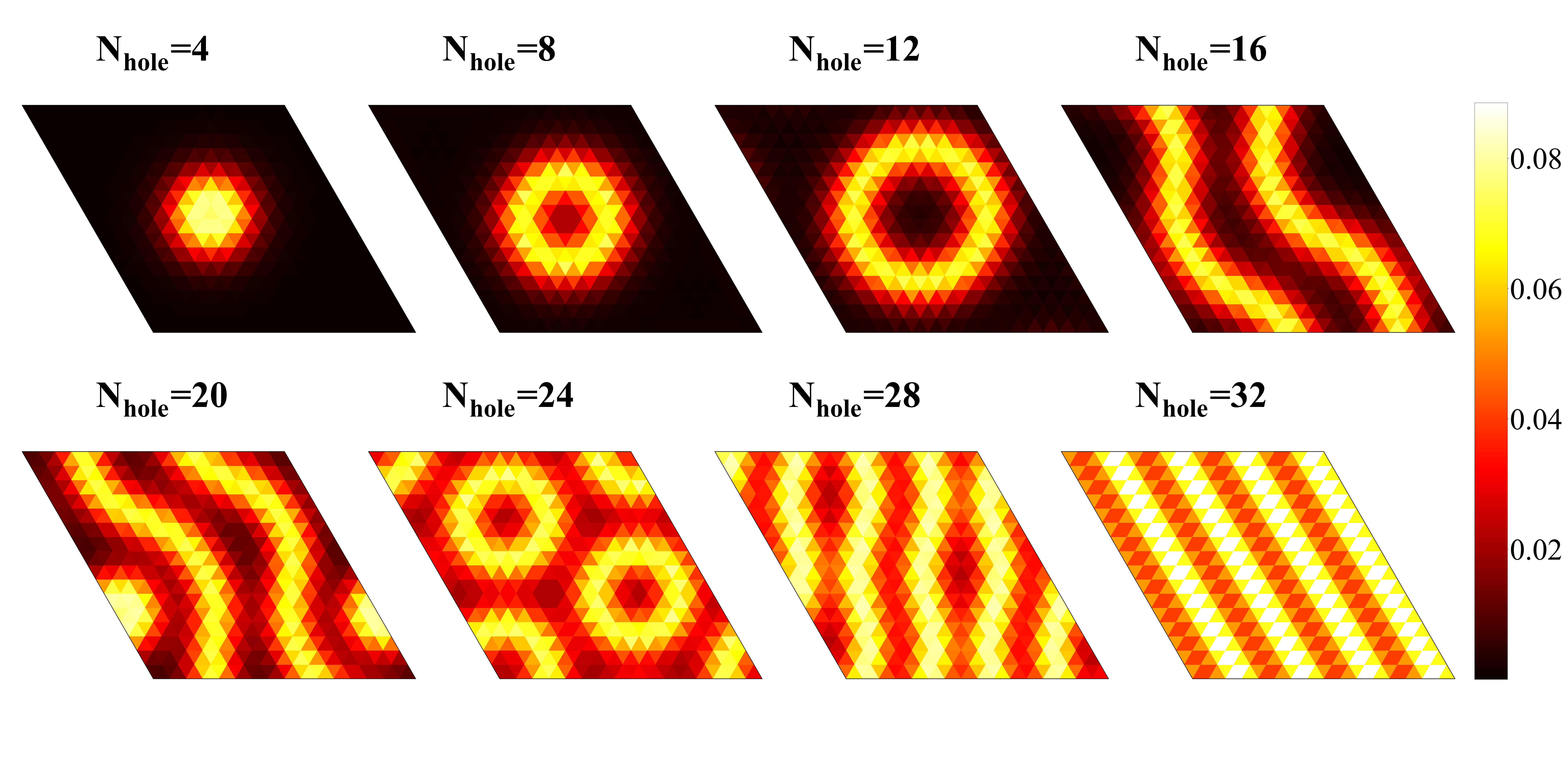}
	\caption{Color map of mean filed results of hole density on a 16$\times$16 system with different dopings and $U = 3$.
	        Half-filled stripe order is developed at $N_{hole} = 32$ which corresponds to $1/16$ doping.}
	\label{f_16_16_h}
\end{figure}

\begin{figure}[t]
	\includegraphics[width=80mm]{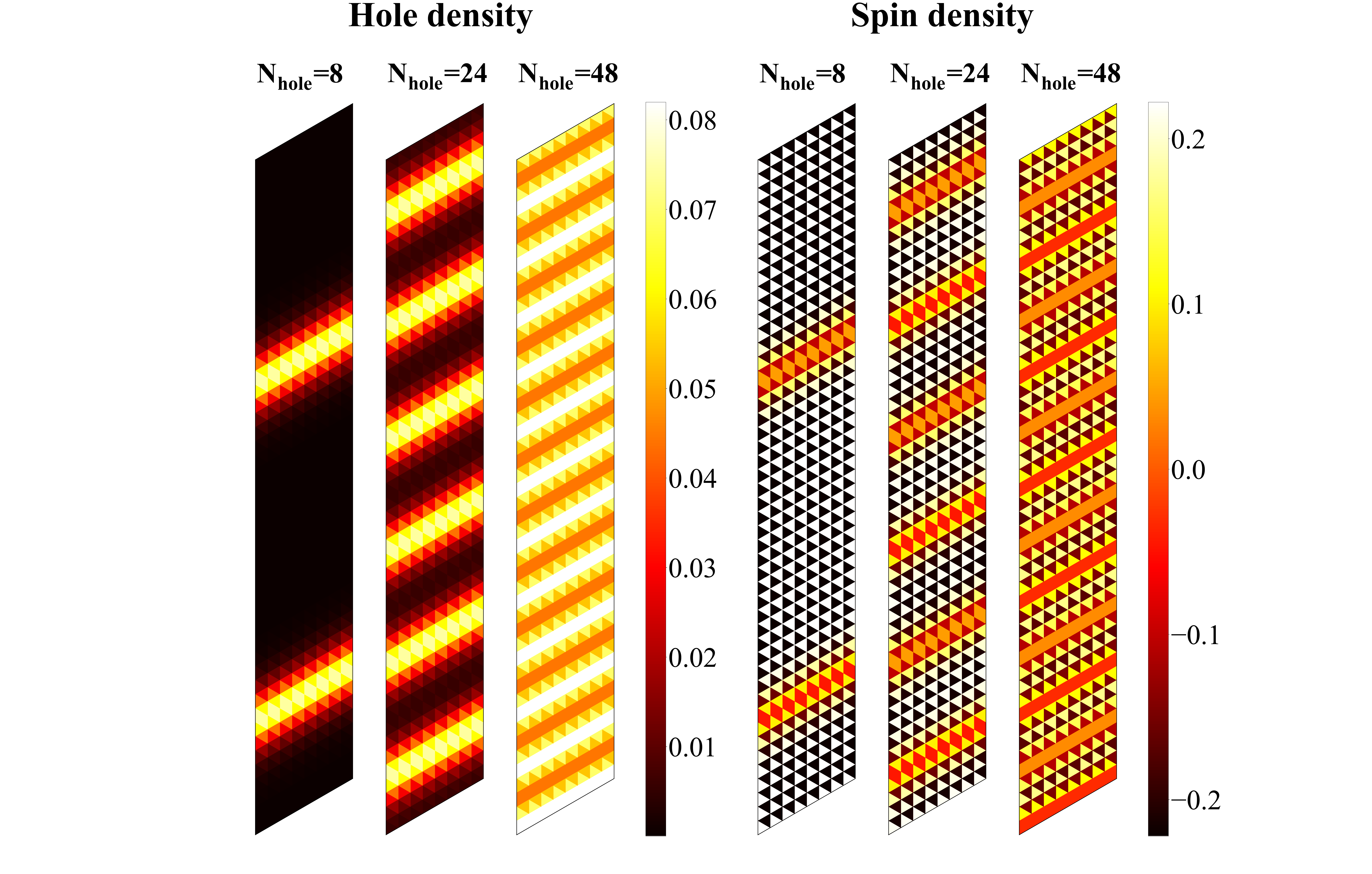}
	\includegraphics[width=80mm]{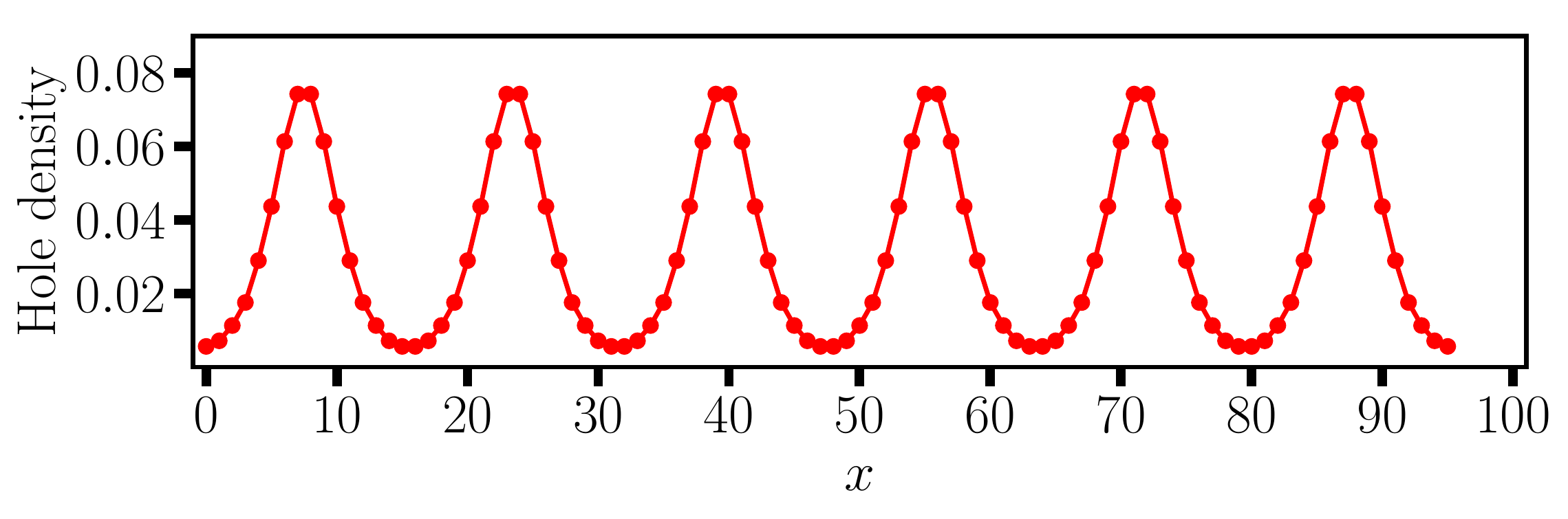}
	\includegraphics[width=80mm]{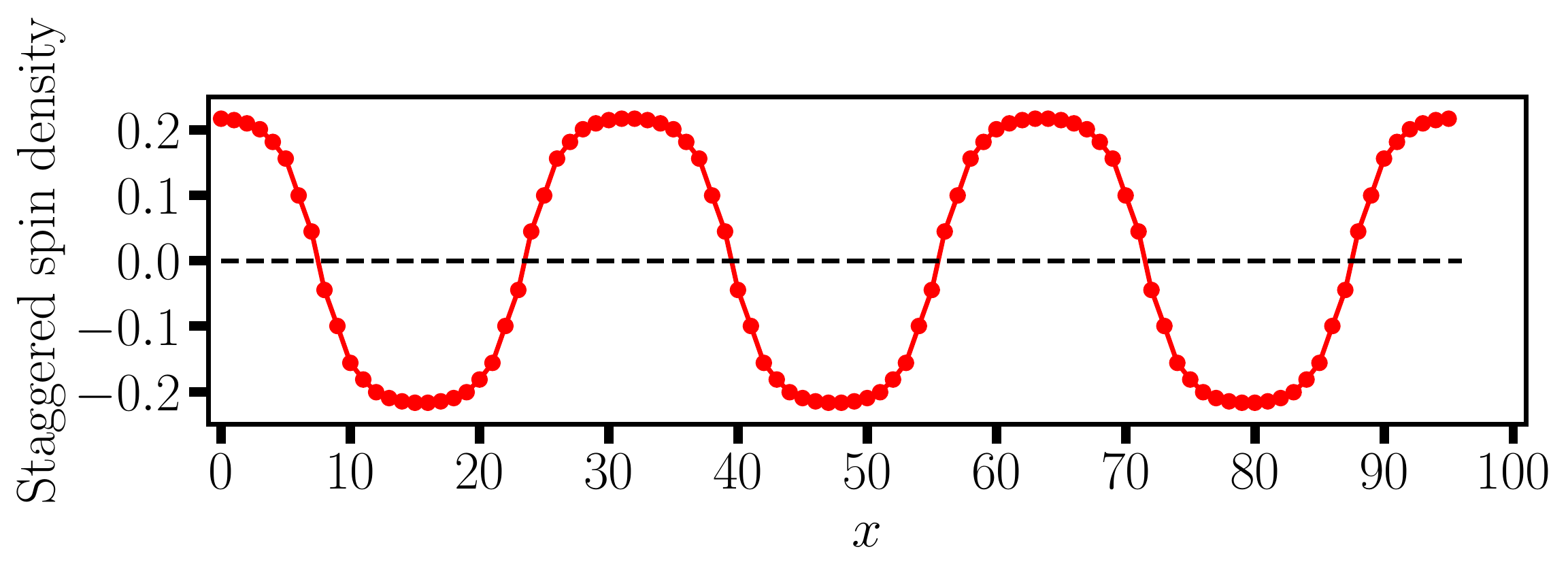}
	\caption{Mean field results for 8$\times$48 system with U=3. Top: color map of hole (left) and spin density (right) for $N_{hole} = 8, 24$, and $48$. Middle (Bottom): plots of hole (staggered spin) density along $L_{2}$ direction with $N_{hole} = 24$.}
	\label{8_48}
\end{figure} 

\subsection{Mean Field}
It is known that for the doped Hubbard model on square lattice, mean-field calculations give stripe order in the ground state
 \cite{PhysRevB.40.7391,doi:10.1143/JPSJ.59.1047,PhysRevB.39.9749,refId0}.
We perform similar mean-field calculations for the honeycomb lattice \cite{Xu_2011}. At half-filling, the
critical interaction strength which separate the semi-metal to the AF insulator phase
is $U_c^{mf} \approx 2.23$ \cite{PhysRevB.101.125103}  (see also in the Appendix \cite{supp}).

Instead of using $U = 8$ as in the DMRG calculation, we choose a renormalized interaction strength with $U \approx 3$ which is above $U_c^{mf} \approx 2.23$, to ensure
the ground state is in the Neel order phase before holes are doped. The effect of interaction strength on the mean-field results is discussed in the Appendix \cite{supp}. In Fig.~\ref{f_16_16_h}, local hole density are plotted on $16 \times 16$ lattices with $U=3$ and for a serial of dopings. With the increase of doping, 
the distribution of the holes in the system changes significantly which was also observed in the square lattice case \cite{Xu_2011}. At low doping, the pattern
shows the symmetry of the lattice. These patterns at low density are basically from finite size effects and they will change with the increase of system size \cite{Xu_2011}. 
 A half-filled stripe order at $N_{hole} = 32$ which corresponding to $1/16$ doping can be found in Fig.~\ref{f_16_16_h}. 


In Fig.~\ref{8_48} we show the mean-field results for the $8 \times 48$ lattice with different number of holes and $U = 3$. 
For $N_{hole} = 8, 24$, and  $48$, states with $2, 6$, and $12$ stripes can be seen from Fig.~\ref{8_48}. Again, all of the stripes are half-filled.
We also show plots for both the local hole and staggered spin density along $L_y$ direction for $N_{hole} = 24$ in Fig.~\ref{8_48}, where the periodic modulation
of them can
be seen. The sharp peak in the hole density plot means holes are very concentrated, while in the DMRG results, hole density shows
a more smooth modulation.
We also notice in the DMRG results for cylinders in Fig.~\ref{DMRG_4_8} and Fig.~\ref{DMRG_4_16}, the hole density at the boundaries is much less than
	the minimum value of the oscillation of hole in the bulk. We also perform a mean-field calculation for the same cylinder in Fig.~\ref{DMRG_4_16} and find
	a similar phenomenon in the mean-field results which indicates open boundary tends to push holes to the bulk. 

 From these results, we know mean-field gives an insulator stripe state where holes are nearly localized in one direction of the lattice,
while in the real many-body ground state as in the DMRG results in Fig.~\ref{DMRG_4_8} and Fig.~\ref{DMRG_4_16}, holes are more dispersed and
is more helpful for the development of superconductivity.  

Other than the half-filled stripe state, we also find stripe state with filling close to $1/2$, $1/3$ and $2/5$ for examples, in the mean-field calculations.
However, the energies of these stripe states are nearly degenerate which indicates the wave-length of the stripe order can fluctuate without causing much
energy as in the square lattice case \cite{Zheng1155}. The details can be found in the Appendix \cite{supp}.

\section{Summary and perspectives}
We discover the half-filled stripe order in the ground state of the doped Hubbard model on honeycomb lattice in the strong 
interaction region by studying width-4 cylinders
with DMRG. We also
carry out complementary mean-field calculations on large systems with renormalized interaction strength and find the half-filled stripe order. 
In the mean-field calculations, the stripe states in the vicinity of half-filling are nearly degenerate in energy, which indicates the wave-length of stripe
can fluctuate freely as in the square lattice case \cite{Zheng1155}. 
Our results indicate that the similarity in real space between the square and honeycomb lattice plays the dominant role in the strong interaction
region given the huge difference between the Fermi-surfaces of 
them. There are real materials \cite{PhysRevB.85.085102} with honeycomb lattice structure, where the stripe order could be 
possibly measured with neutron scattering \cite{Tranquada-2004} or
scanning tunneling microscope \cite{Kohsaka1380} techniques. The stripe order could also be observed on artificially synthesized systems with honeycomb structure \cite{2013NatNa...8..625P}.
The role of stripe order to the 
possible superconducting pairing order of the doped Hubbard model on honeycomb lattice is a question one may ask. 
The presence of stripe order means in the calculation of pairing order or correlation, unit cell compatible to the stripe order needs to be chosen
in order not to frustrate the stripe order. How the stripe order melts with thermal
fluctuation \cite{2020arXiv200910736W} is an interesting question since experiments are performed at 
finite temperature. Whether stripe order is a universal consequence
of doping anti-ferromagnetic Mott insulator is an interesting topic for future investigation.

\begin{acknowledgments}
M. Q. thank C.-M. Chung, Schollw\"{o}ck, S. R. White and S. Zhang for earlier collaborations on related topic.
X.Y. and H. Z. acknowledge the financial support from National Natural Science Foundation of China ((Grants No. 92065201, Grants No. 11790313, No. 11674226, No. 11861161003, No. 11521404, No. 11634009, No. 11674222, No. 11874256, No. U1632102, and No. 11874258), National Key Research and Development Program of China (Grants No. 2016YFA0300403, No. 2016YFA0301003), Science and Technology Commission of Shanghai Municipality (Grants No. 19JC1412701, No. 2019SHZDZX01).
M. Q. is supported by a start-up fund from School of Physics and Astronomy in Shanghai Jiao Tong University.
M. Q. also acknowledge the support of computational resources by S. Zhang at the Flatiron Institute. The DMRG calculations
in this work are performed with the iTensor package \cite{itensor}.
\end{acknowledgments}

\bibliography{Honeycomb.bib}

\begin{thebibliography}{65}
\bibitem[{foo({\natexlab{a}})}]{foot1}
\bibinfo{note}{We also perform mean-field calculation for a width-4 cylinder to
  directly compare with DMRG results.}
\end{thebibliography}

\appendix

\section{The convergence of DMRG results}

\begin{figure}[t]
	\includegraphics[width=80mm]{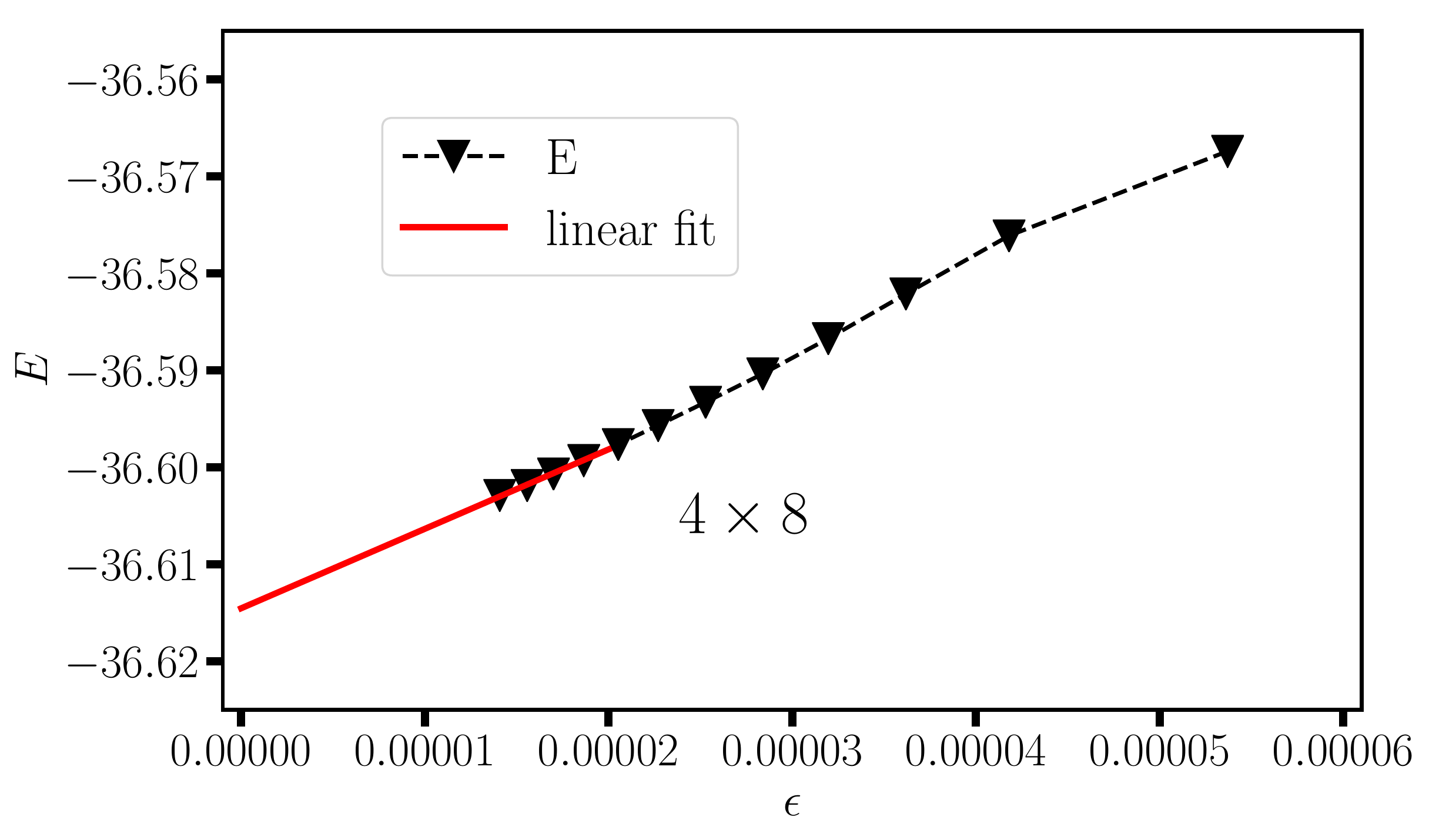}
	\includegraphics[width=80mm]{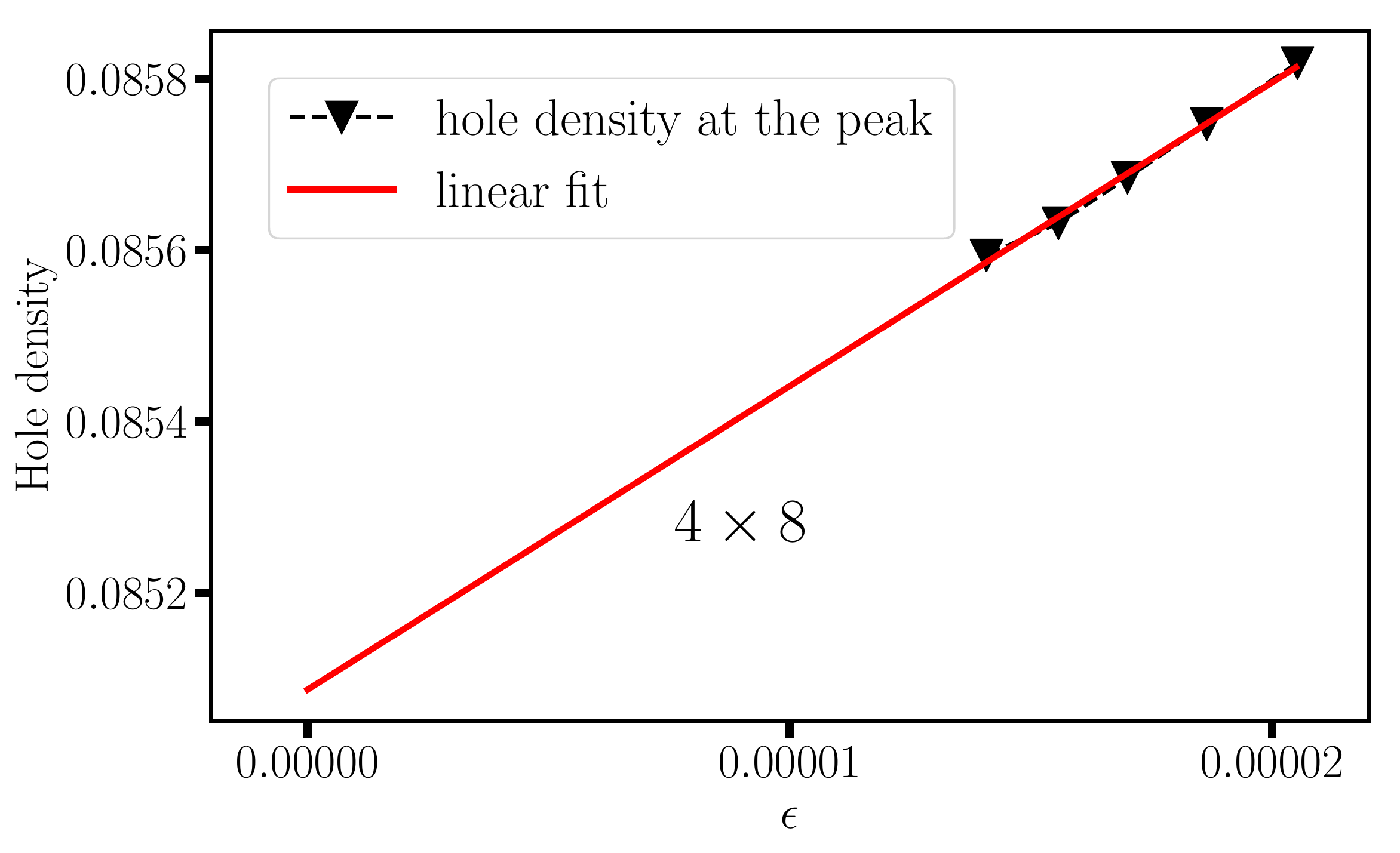}
	\caption{Scaling of energy and the peak value of hole density versus the truncation error in DMRG for the $4 \times 8$ cylinder. The fit is performed using the
		last $5$ points. The
		largest state kept $M$ is $9000$ (the last point).
		Upper: energy.  Lower: peak hole density.}
	\label{E_scaling_4_8}
\end{figure} 

\begin{figure}[t]
	\includegraphics[width=80mm]{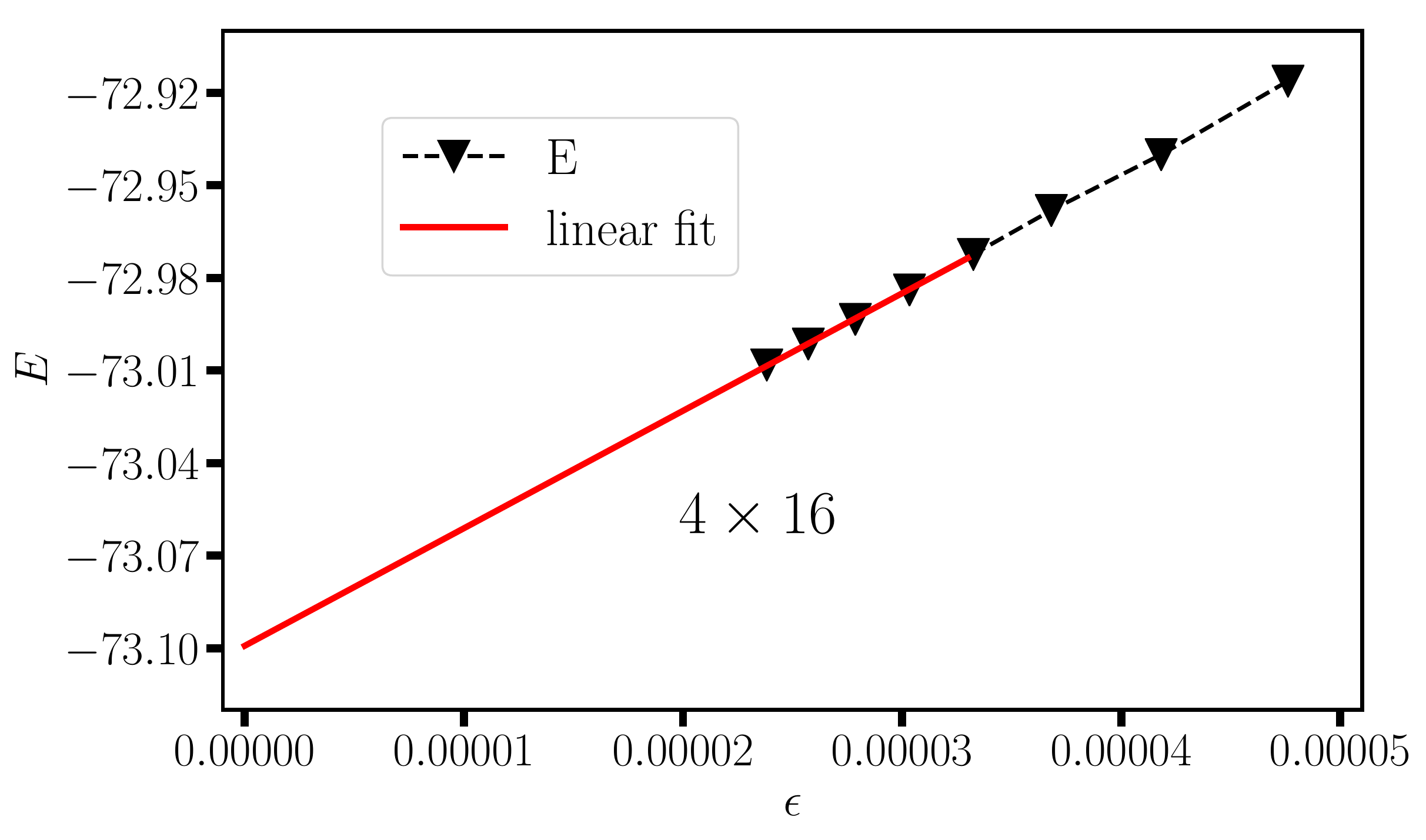}
	\includegraphics[width=80mm]{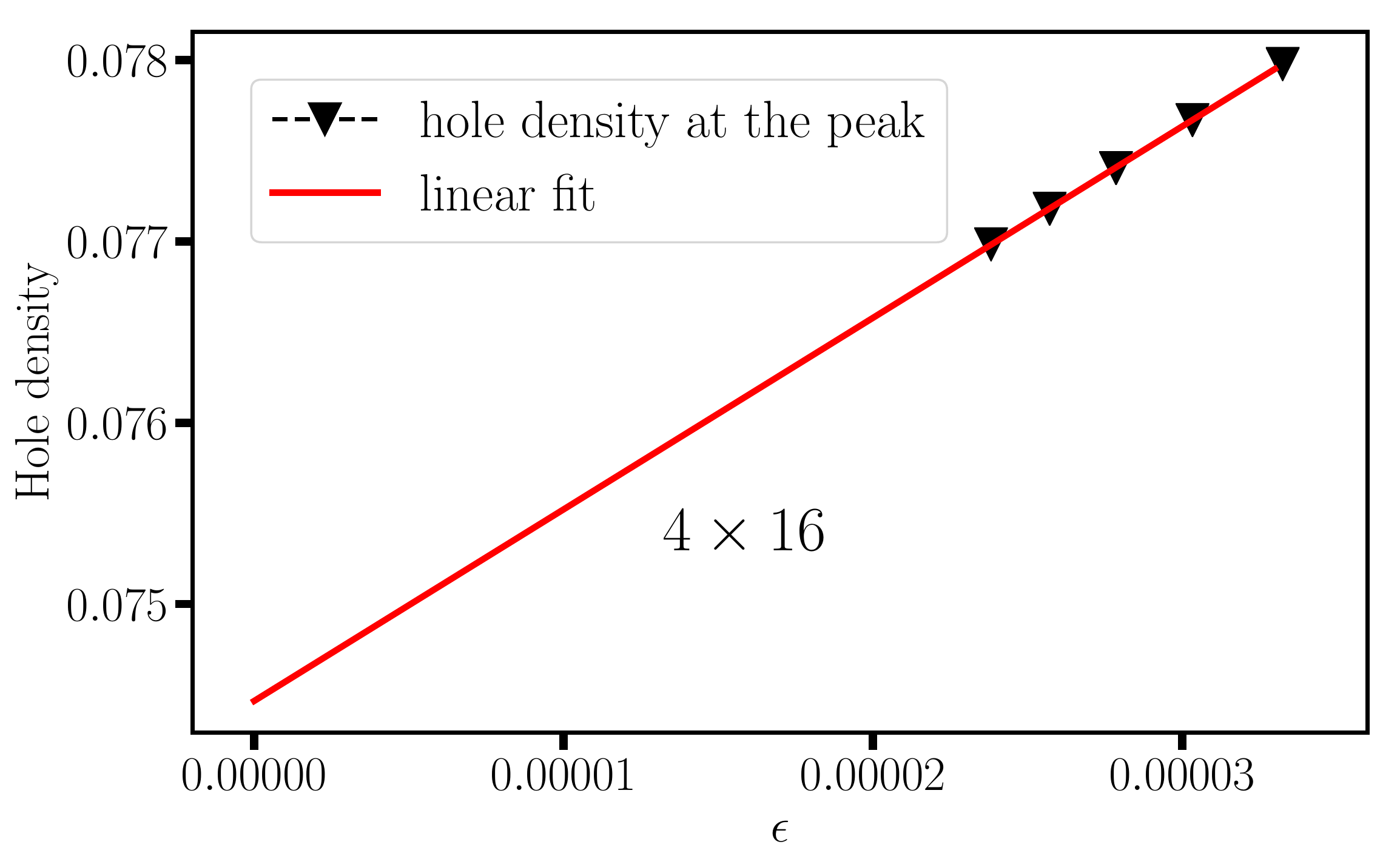}
	\caption{Scaling of energy and the peak value of hole density versus the truncation error in DMRG for $4 \times 16$ cylinder. The fit is performed using the last $5$ points. The
		largest state kept $M$ is $7000$ (the last point).
		Upper: energy.  Lower: peak hole density.}
	\label{E_scaling_4_16}
\end{figure}

\begin{figure}[t]
	\includegraphics[width=80mm]{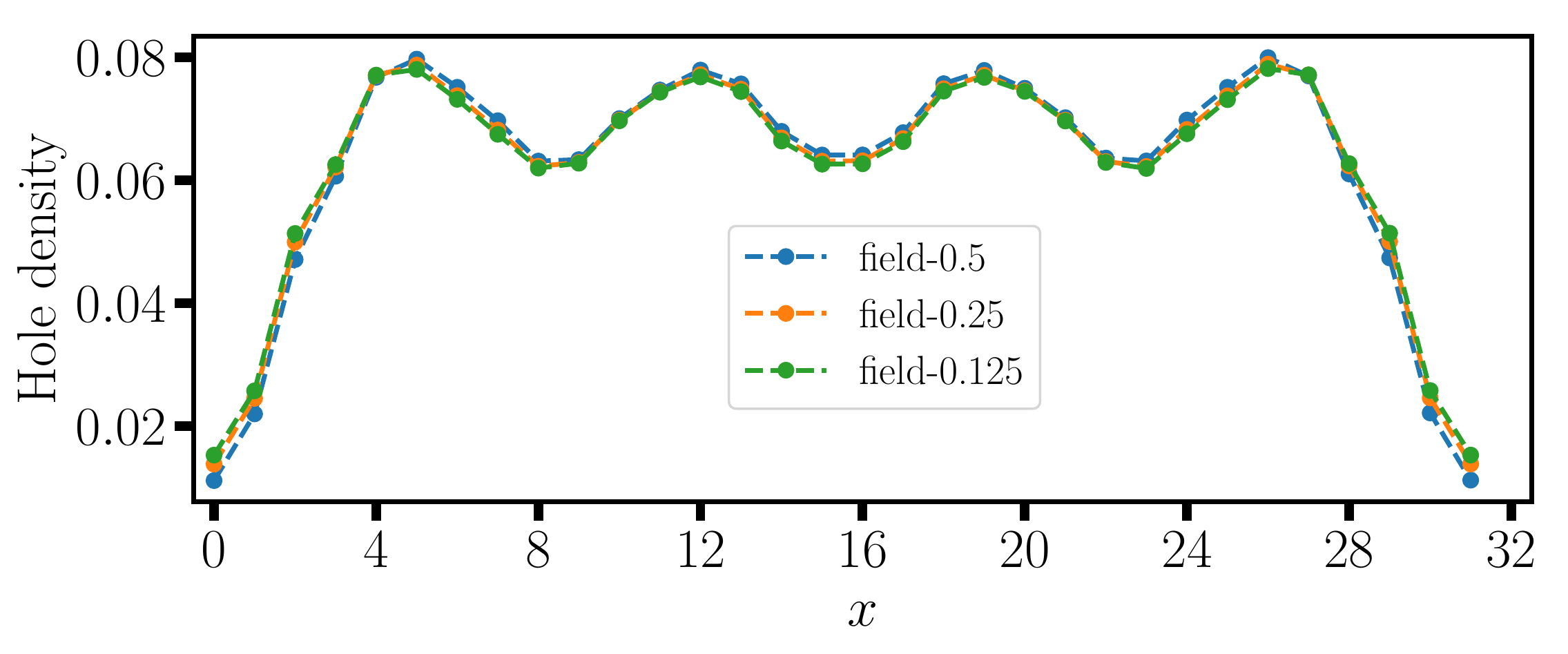}
	\includegraphics[width=80mm]{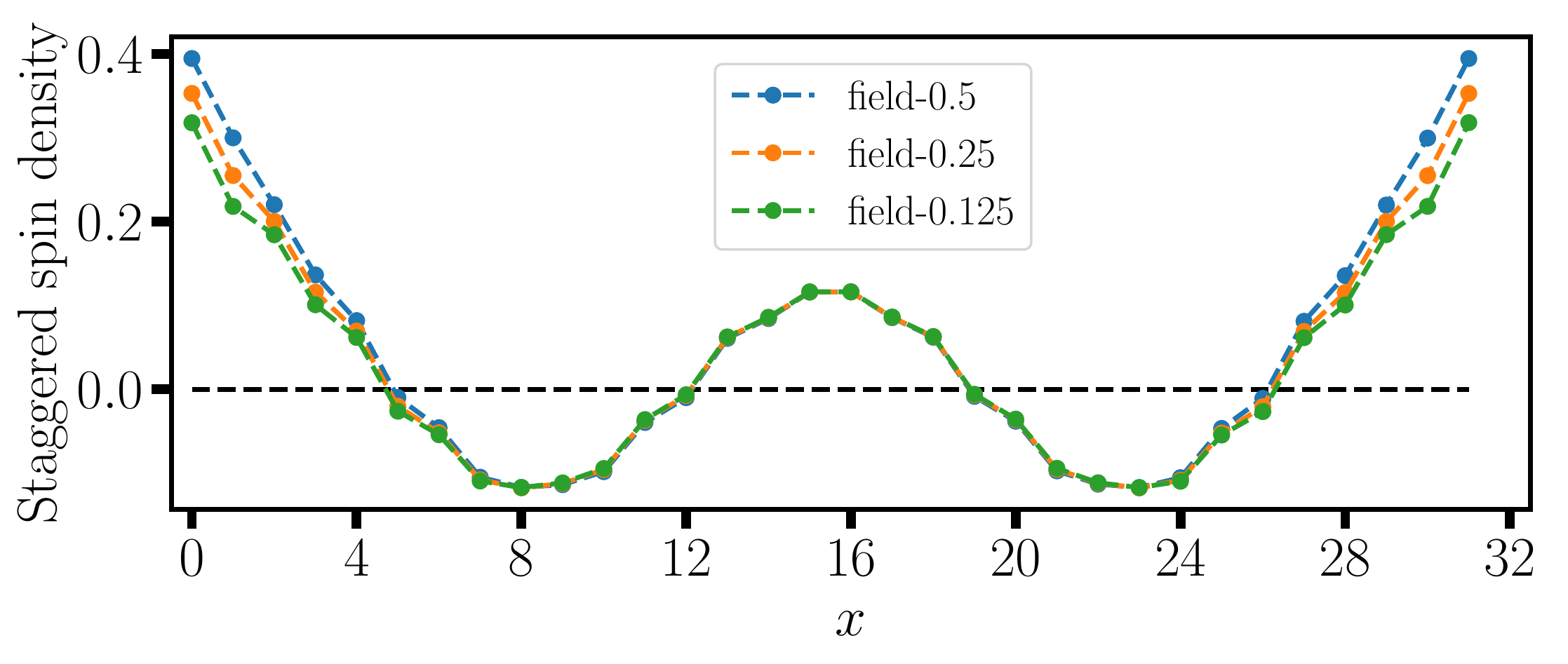}
	\caption{{DMRG results of the hole and spin density for a $4 \times 16$ cylinder with $U = 8$ and $8$ holes.
			Upper (Lower): plots of hole (staggered spin) density along $L_{2}$ direction with kept state $m = 5000$.
			Results for boundary pinning fields with strengths $h_p = 0.5, 0.25$, and $0.125$ are shown. We can find the strength
			of pinning fields only affects the results near the boundaries.}}
	\label{DMRG_4_16_boundary}
\end{figure}

\begin{figure}[b]
	\includegraphics[width=40mm]{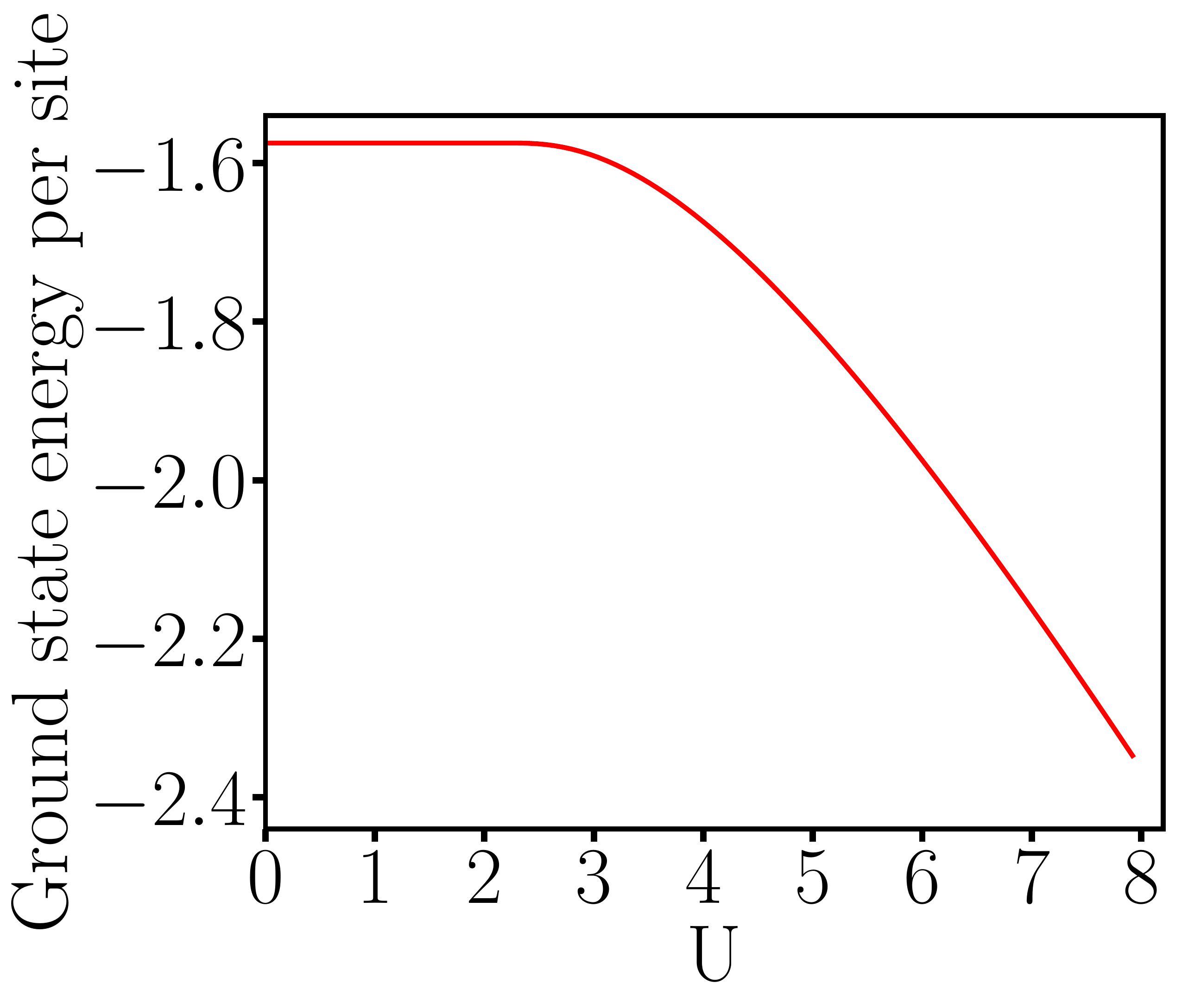}
	\includegraphics[width=40mm]{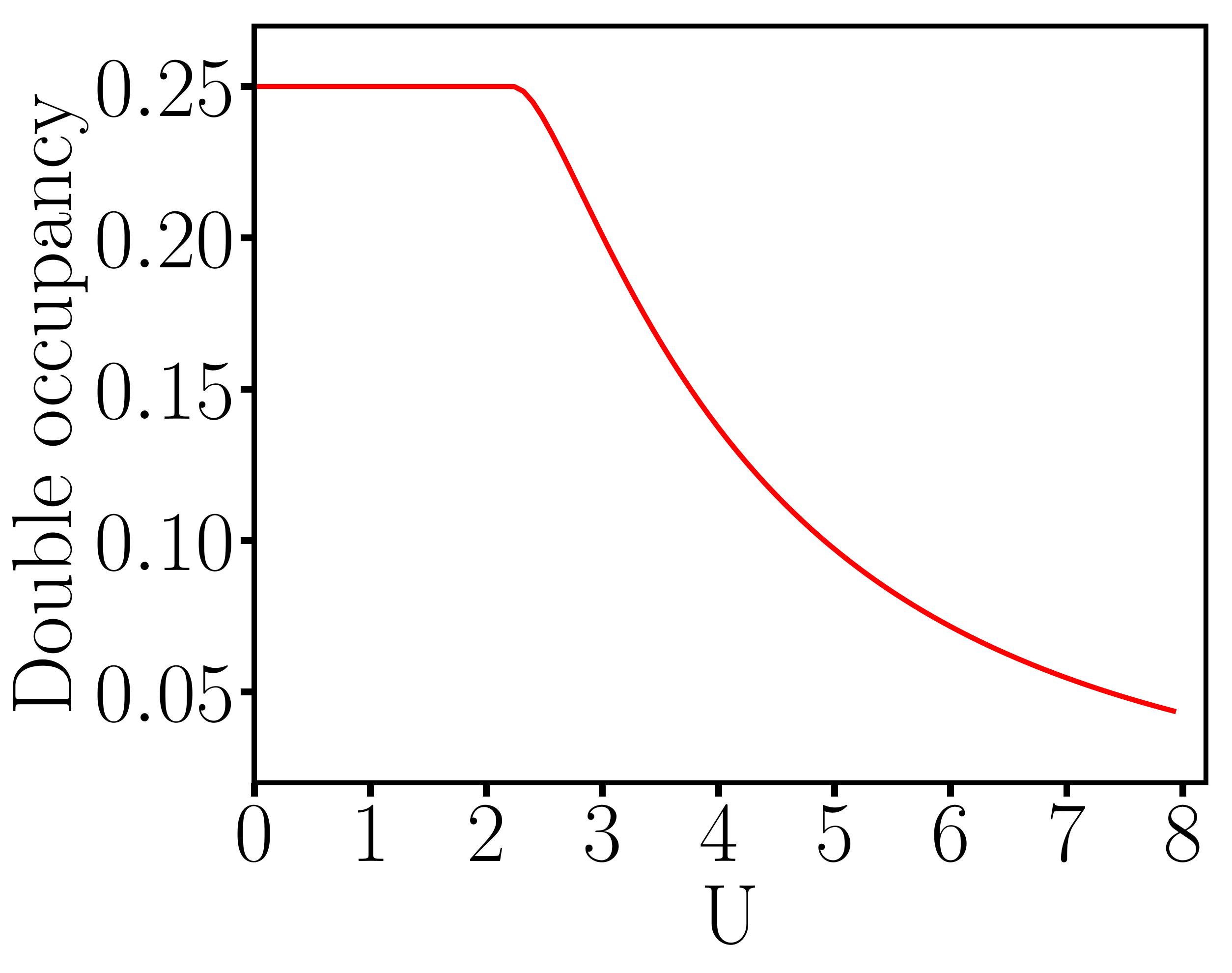}
	\includegraphics[width=40mm]{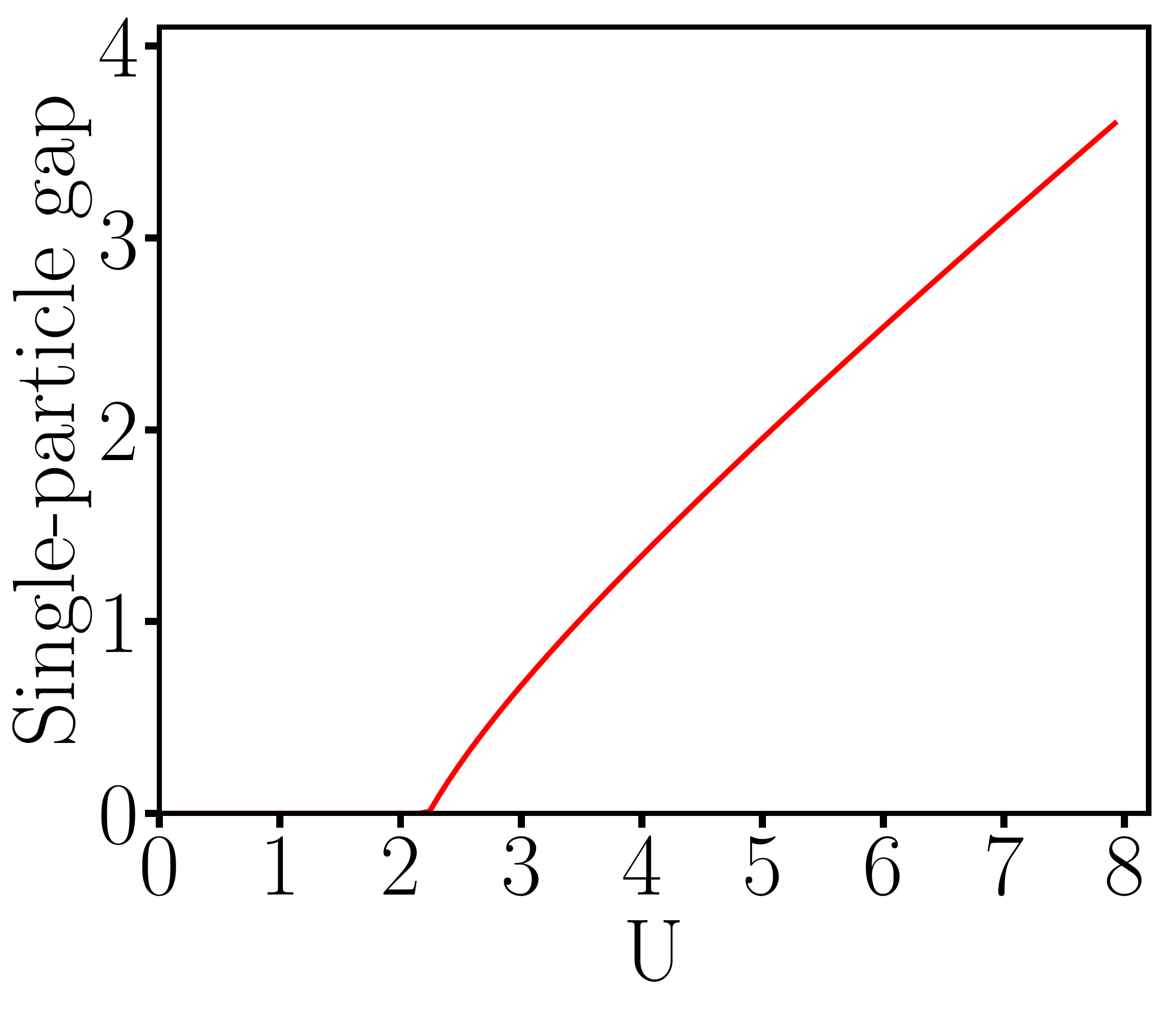}
	\includegraphics[width=40mm]{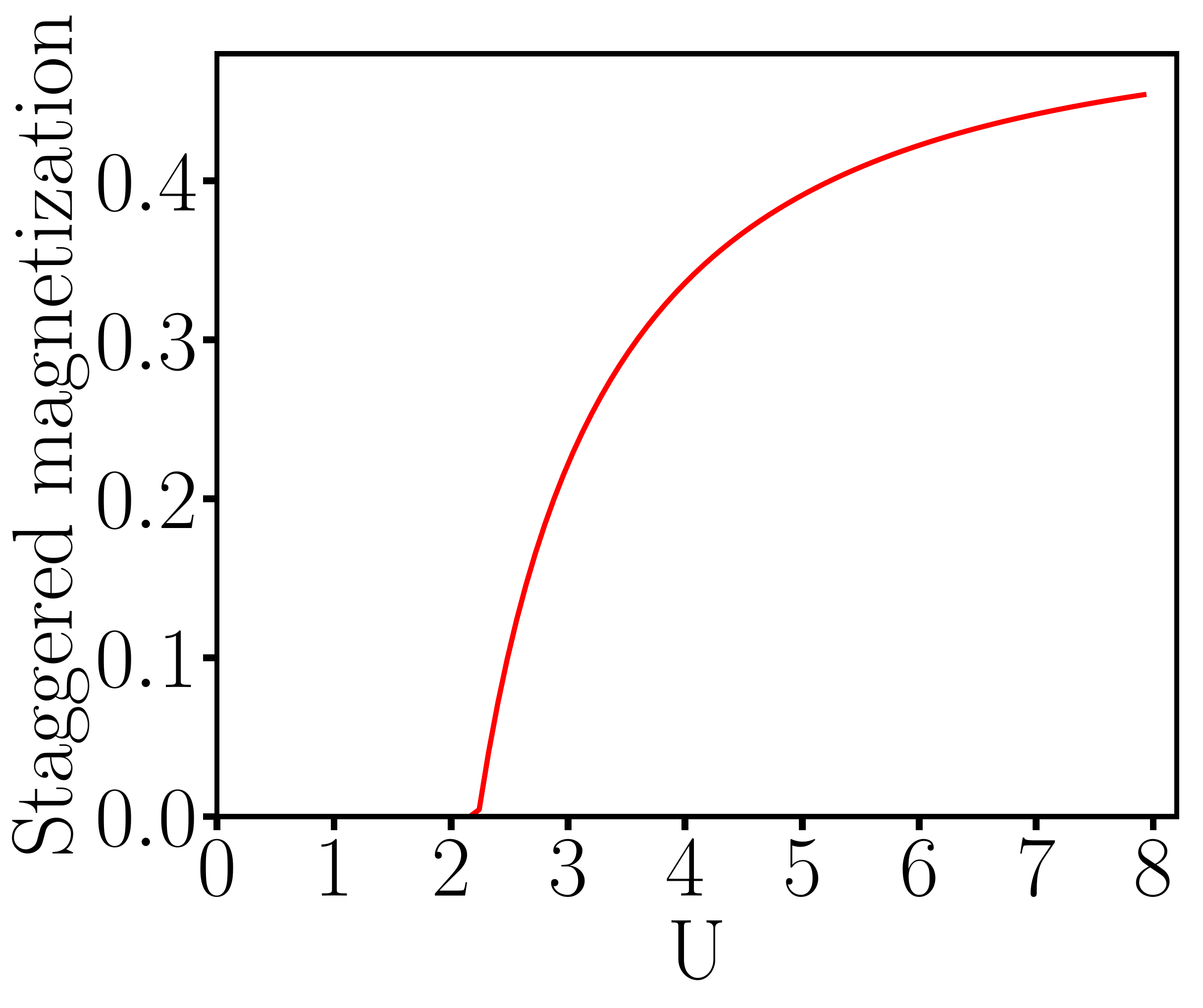}
	\caption{Mean-filed results for ground state energy per site, double occupancy, single particle gap, and staggered magnetization at half filling in the thermaldynamic limit.
		We can see a metal-insulator transition at $U \approx 2.23$.}
	\label{mf-half}
\end{figure}

\begin{figure}[t]
	\includegraphics[width=80mm]{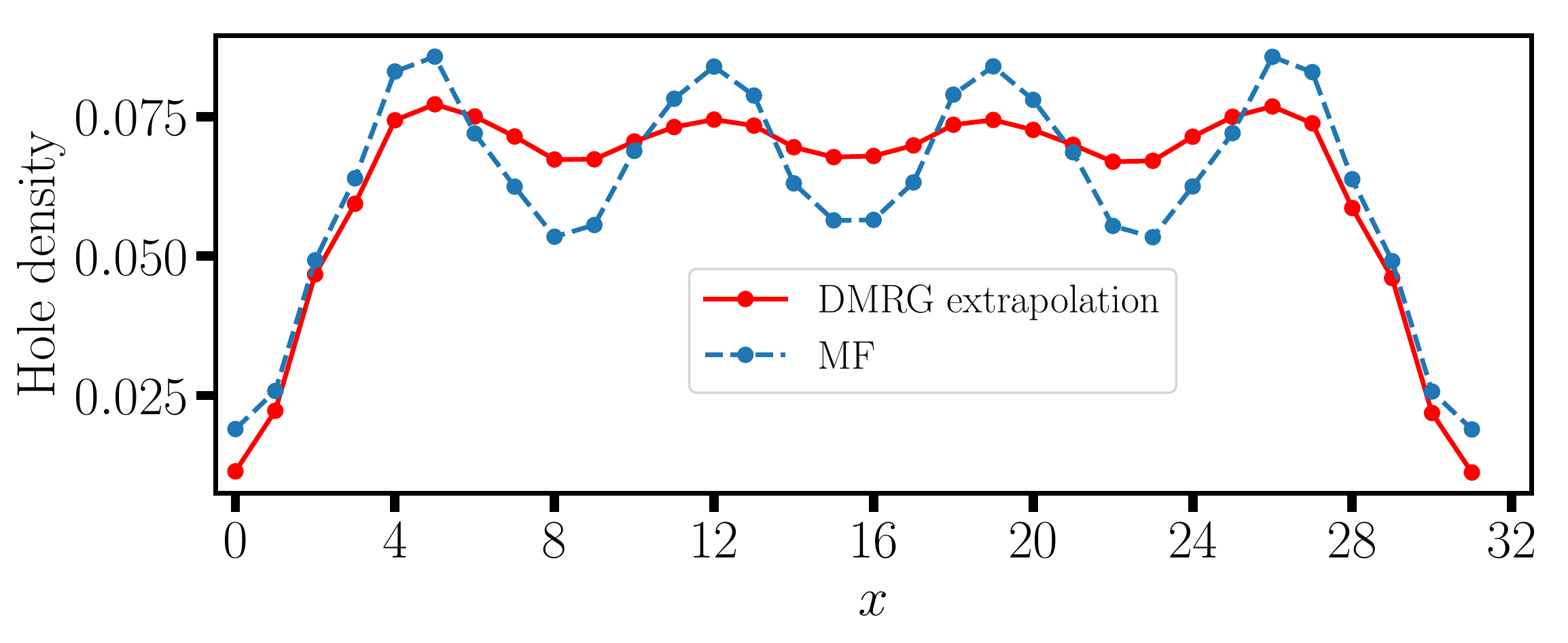}
	\includegraphics[width=80mm]{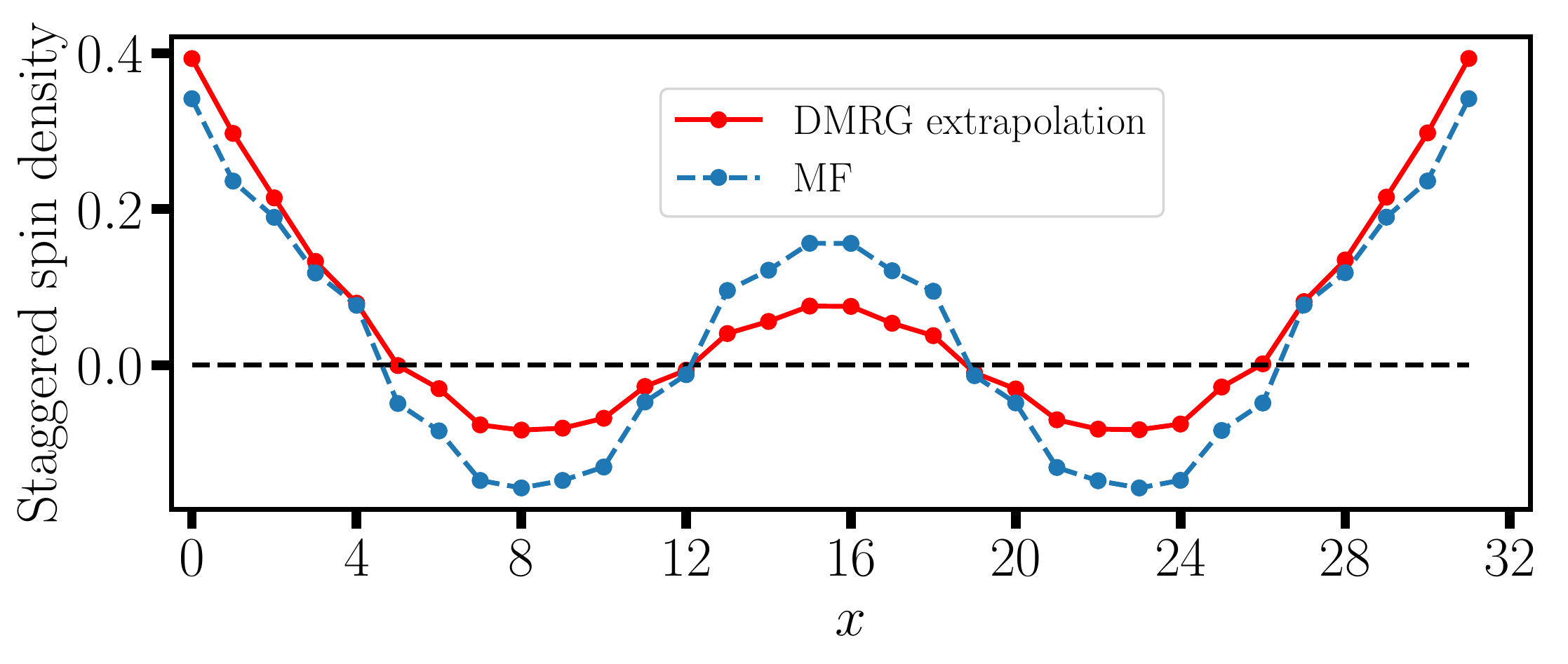}
	\caption{{ Extrapolated DMRG and mean-field results of the hole and spin density for a $4 \times 16$ cylinder with $8$ holes.
			Pinning fields with strength $h_p = 0.5$ are applied at the boundaries.
			The DMRG results are for $U = 8$ while the mean-field results are with an effective $U = 3$.
			Upper (Lower): plots of hole (staggered spin) density along $L_{2}$ direction. We can see even with a smaller effective $U$
			value, the stripe order in mean-field is larger than DMRG results. However, the stripe structure is the same.
	}}
	\label{DMRG_4_16_mf}
\end{figure}

\begin{figure}[t]
	\includegraphics[width=80mm]{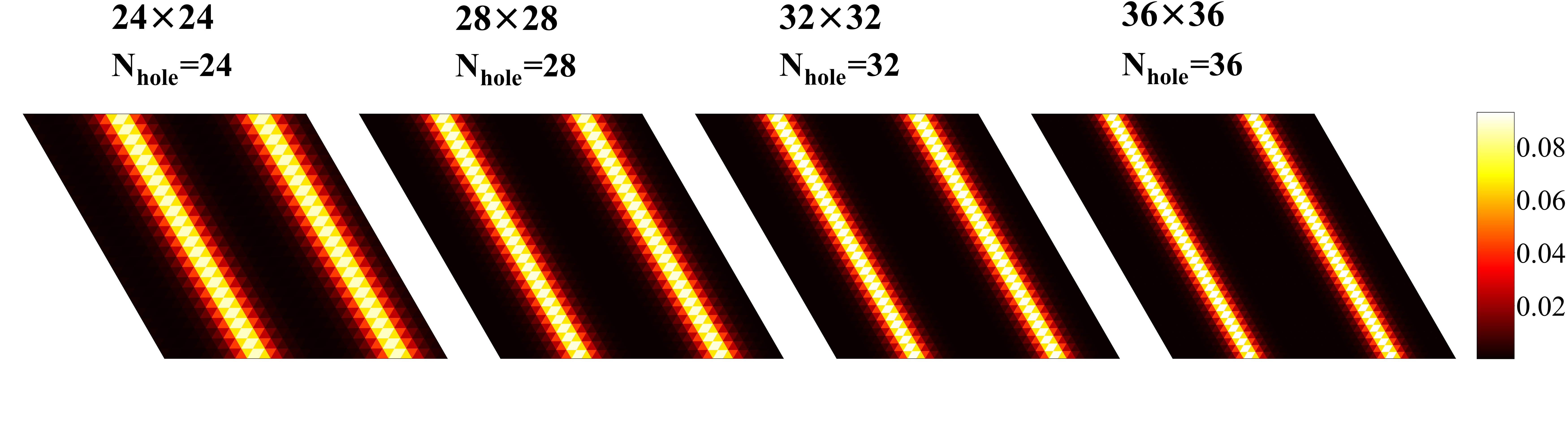}
	\caption{Color map of hole density on a series of lattices with sizes $24 \times 24$, $28 \times 28$, $32 \times32$, and $36 \times 36$ at $U=3.16$, and with
		$N_{hole} = 24,28,32$, and $36$ respectively.}
	\label{compare_1_4h}
\end{figure} 

In this section, we check the convergence of results with the kept state $m$ in the DMRG
calculations. In Fig.~\ref{E_scaling_4_8} and Fig.~\ref{E_scaling_4_16}, we plot the scaling of energy and the peak value of hole density versus the truncation error for
$4 \times 8$ and $4 \times 16$ systems respectively. A linear fit using the five points with the
largest $m$ (smallest truncation error $\epsilon$) is performed. The largest $m$ are $9000$
and $7000$ for $4 \times 8$ and $4 \times 16$ respectively. As we can see from Fig.~\ref{E_scaling_4_8} and Fig.~\ref{E_scaling_4_16},
both energy and peak value of hole density reach the linear scaling region in our calculation, which ensures the
reliability of the extrapolation to zero truncation error.

{
	\section{The boundary effects}
	We study the effect of the strength of pinning fields at boundaries for a $4 \times 16 $ cylinder with fixed kept state $m = 5000$ in Fig.~\ref*{DMRG_4_16_boundary}.
	As we can see in Fig.~\ref*{DMRG_4_16_boundary}, by decreasing the strength of pinning fields, the hole density remains nearly unchanged. The spin density near the
	boundaries decrease while the ``bulk" value doesn't change. This result indicates the stripe order in the bulk is robust against boundary effect. 
}

\section{Mean-field calculation}

We break the Hamiltonian of Hubbard model in the following form in the mean-field calculation:
\begin{equation}
\begin{split}
H^{MF}=&-t\sum_{\langle i,j\rangle,\sigma}c_{i\sigma}^\dagger c_{j\sigma}\\
&+U\sum_{i}(\langle n_{i\uparrow}\rangle n_{i\downarrow}+n_{i\uparrow}\langle n_{i\downarrow}\rangle -\langle n_{i\uparrow}\rangle \langle n_{i\downarrow}\rangle)
\end{split}
\label{mf_H}
\end{equation}
where $\langle n_{i\sigma}\rangle(\sigma=\uparrow,\downarrow)$ is the density on site $i$ which needs to be determined self-consistently.
Because spin degrees of freedom are decoupled, the mean-field Hamiltonian in Eq.~(\ref{mf_H}) is
block-diagonal:
\[ \mathbf{H^{MF}} = \left(
\begin{array}{cc}
H^{MF}_{\uparrow} &  \\
& H^{MF}_{\downarrow} \\
\end{array} \right) \]
Hence we can diagonalize the block matrix $H^{MF}_{\sigma}(\sigma=\uparrow,\downarrow)$ independently
for given $\langle n_{i\sigma}\rangle(\sigma=\uparrow,\downarrow)$.
The ground state is determined by solving the Hamiltonian self-consistently.

We first randomly select initial values for $\langle n_{i\sigma}\rangle(\sigma=\uparrow,\downarrow)$.
Then we diagonalize the Hamiltonian in Eq.~(\ref{mf_H}), from which a wavefucntion of $N_{\sigma}$ particles are obtained.
We can calculate the local density from this wavefucntion, which is then used to update the Hamiltonian. This process is repeated untill the density is convergent.

The convergence is not guaranteed if the density calculated from last step wave-function is taken as input for the current step.
To improve convergence, we take advantage of a technique called Anderson mixing \cite{Anderson1965Iterative}, which has widely usage in electronic structure computations.
The wave-function could converge to a local minimum, if we start from random initial densities. To make sure the global minimum is reached, additional strategies are adopted.
We try different initial densities and the global minimum is determined as the one with lowest energy. We also add small perturbation to the converged wave-function, and then
use it as new initial densities for the self-consistent iteration. This procedure is repeated several times. 

We can also test whether a certain type of stripe state is the ground state by using it as the initial wave-function.
After we obtain a stripe state on small lattices, we can construct a wave-function on large lattice to test whether it is 
also the solution for large system.

\subsection{Half-filling case}
In Fig.~\ref{mf-half} we show the mean field results for ground state energy, double occupancy, single particle gap, and AF magnetization at half filling in the thermaldynamic limit.
From Fig.~\ref{mf-half}, we can see a second-order metal-insulator transition occurs at $U \approx 2.23$ at half-filling.

\subsection{Comparison with DMRG results}
{
	In Fig.~\ref*{DMRG_4_16_mf} we show a comparison of the DMRG (extrapolated to zero truncation error) and mean-field results for a $4 \times 16$ cylinder at $1/16$ doping
	with pining field $h_m = 0.5$ applied 
	at the boundaries. The DMRG results are for $U = 8$ while a reduced effective interaction $U = 3$ is used in the mean-field calculation. From Fig.~\ref*{DMRG_4_16_mf} we can see
	the stripe order from DMRG is weaker than the mean-field results, which is a manifestation of the quantum fluctuation. However, from the DMRG results we can conclude
	the stripe order is not killed by quantum fluctuation. We also notice in both results, the hole density at boundaries is much less than the minimum value in the bulk,
	which means the open boundaries have a effect of pushing holes to the bulk and increase the average hole densities in the bulk for finite system.}

\subsection{The effect of U}

%
%

\begin{figure}[t]
	\includegraphics[width=80mm]{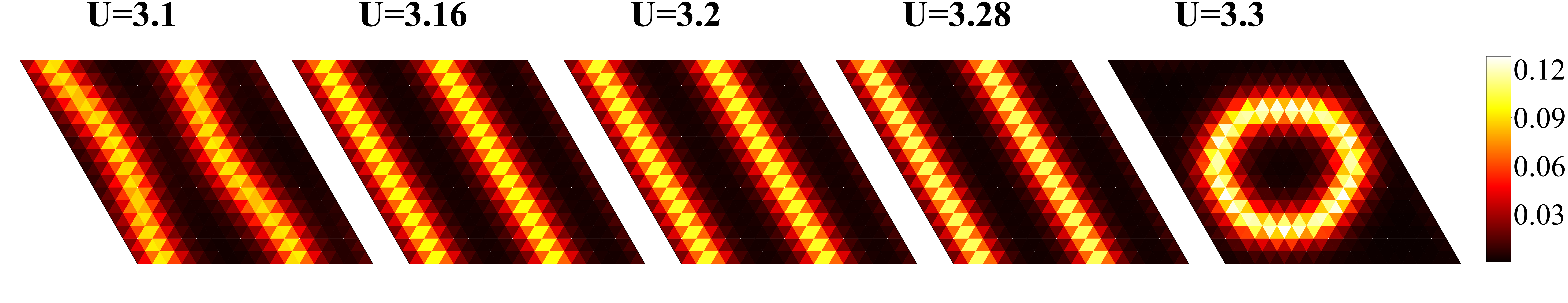}
	\caption{Color map of hole density on a $16 \times 16$ lattice with $N_{hole} = 16$ and $U = 3.1, 3.16, 3.2, 3.28$, and $3.3$.}
	\label{f_16_16_Nhole160}
\end{figure}

In Fig.~\ref{compare_1_4h} we show the ground states for large system sizes from mean-field calculations. As we can see, all
of them are half-filled stripe states. The stripe state can be only obtained with $U>3$. 
Fig.~\ref{f_16_16_Nhole160} shows the effect of $U$ on 16$\times$16 lattice. As we can see, stripe states is only found
for U ranging from 3.1 to 3.3, which also explains why we do not find stripe at $N_{hole} = 16$ in Fig.~4 in the main text. 

%

\subsection{Stripe states with different fillings}

\begin{figure}[t]
	\includegraphics[width=80mm]{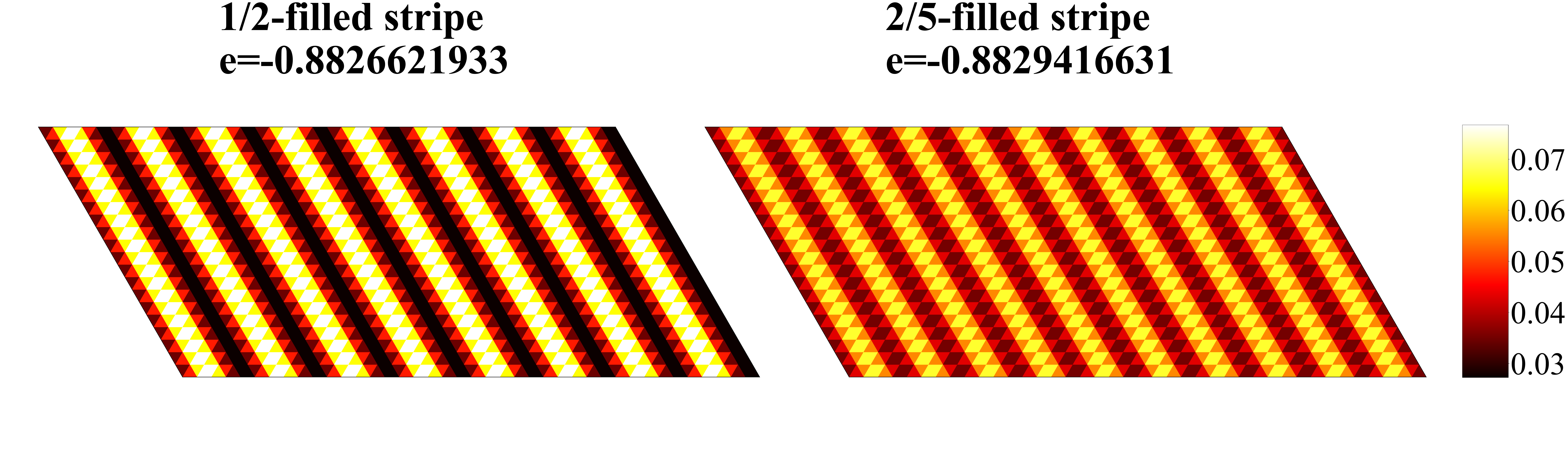}
	\includegraphics[width=80mm]{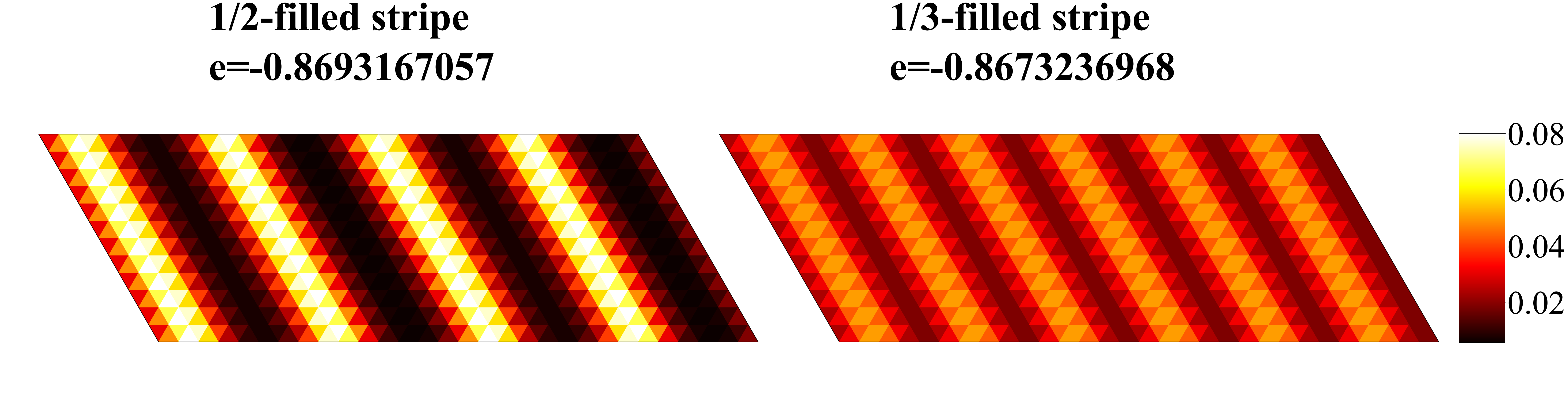}
	\includegraphics[width=80mm]{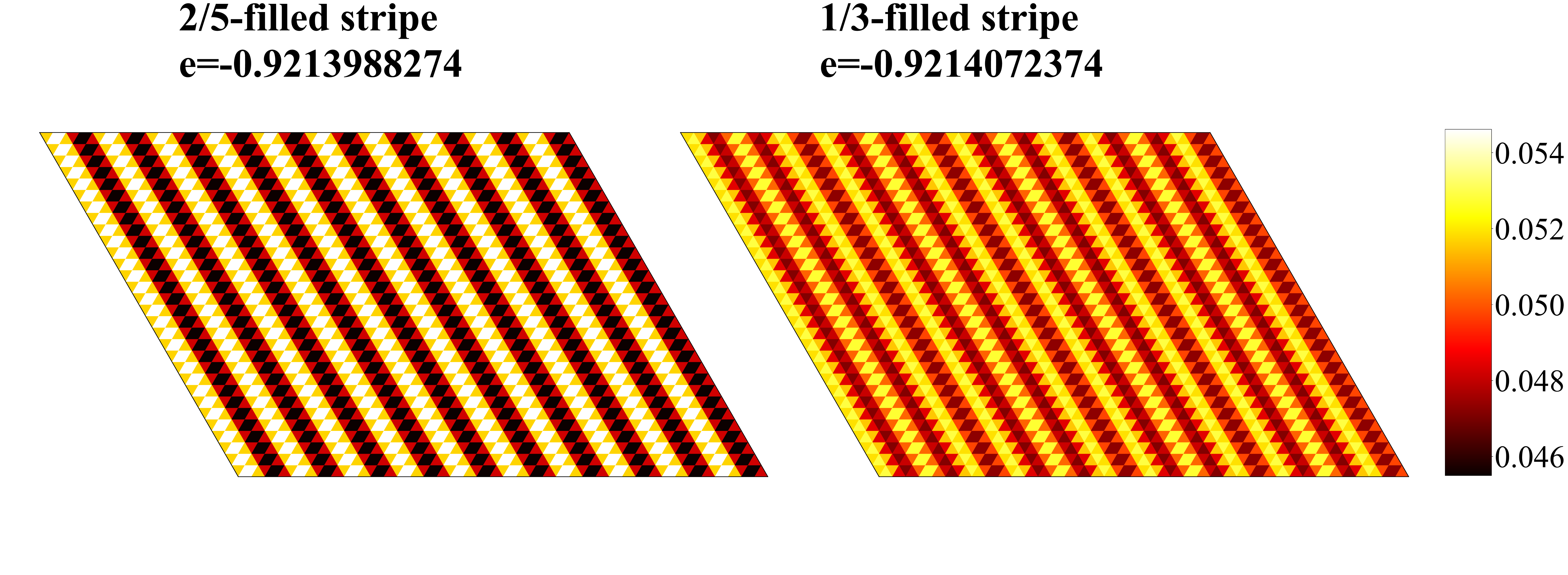}
	\caption{Comparison of energies for stripe states with different fillings.
		Top: color map of hole density on a $20 \times 40$ lattice with $U=3$ and $N_{hole}=80$.
		The filling of stripes are $2/5$ and $1/2$ for left and right states. 
		Middle:  $12 \times 30$ lattice with $U=3$ and $N_{hole}=24$.
		The filling of stripes are $1/2$ and $1/3$ for left and right states.
		Bottom: $30 \times 40$ lattice with $U=2.82$ and $N_{hole}=120$.
		The filling of stripes are $2/5$ and $1/3$ for left and right states.
		In all the cases, both of the stripe states are stable in the mean-filed calculation and the energies for them are almost degenerate.}
	\label{f_40_20_h}
\end{figure}

In Fig.~\ref{f_40_20_h}, we study lattice with size $20 \times 40$ compatible for stripe states with both $1/2$ and $2/5$ filling, 
lattice with size $12 \times 30$ compatible for stripe states with both $1/2$ and $1/3$ filling,
and lattice with size $30 \times 40$ compatible for stripe states with both $2/5$ and $1/3$ filling.
In all the cases, both stripes state are stable in the mean-filed calculation and we find they are almost degenerate in energy.


\end{document}